\documentclass{article}
\usepackage{arxiv}

\usepackage[utf8]{inputenc} 
\usepackage[T1]{fontenc}    
\usepackage{hyperref}       
\usepackage{url}            
\usepackage{booktabs}       
\usepackage{amsfonts}       
\usepackage{nicefrac}       
\usepackage{microtype}      
\usepackage{lipsum}		
\usepackage{graphicx}

\usepackage{array}
\usepackage{amssymb, amsmath, amsthm}
\usepackage{graphicx}
\usepackage{lmodern,url}
\usepackage{cite}
\usepackage{makecell} 
\usepackage{cancel}
\usepackage{multirow}
\usepackage{microtype}
\usepackage{lineno}
\usepackage{lscape}
\usepackage{xspace}
\usepackage{xcolor}
\usepackage{siunitx}
\usepackage{todonotes}
\usepackage{booktabs}
\usepackage[ruled,vlined]{algorithm2e}
\usepackage{optidef}
\usepackage{mathdots}
\usepackage{subcaption}
\usepackage{csquotes}
\usepackage{caption}
\usepackage{verbatim}
\usepackage{ulem}
\graphicspath{{Figures/}}

\usepackage{amsmath}
\usepackage{tikz}
\usetikzlibrary{fit}
\usetikzlibrary{positioning}
\tikzset{%
  highlight/.style={rectangle,rounded corners,fill=red!15,draw,fill opacity=0.25,thick,inner sep=0pt}
}

\definecolor{mygreen}{RGB}{160, 242,182}

\definecolor{mypurple}{RGB}{236, 223, 234}
\newcommand{\PrEP}{\ensuremath{P}\xspace}

\newcommand{\Iasti}{\ensuremath{I^{a}}\xspace}
\newcommand{\Issti}{\ensuremath{I^{s}}\xspace}
\newcommand{\Ssti}{\ensuremath{S}\xspace}
\newcommand{\Tsti}{\ensuremath{T}\xspace}

\allowdisplaybreaks

\title{Stability and bifurcations of a minimal model for the effect of PrEP-related risk compensation in epidemics of sexually transmitted infections}

\usepackage{authblk}

\author[1,2$\dagger$]{Piklu Mallick}
\author[1,2$\dagger$]{Laura Müller}
\author[1,2$\dagger$]{Antonio B. Marín-Carballo}
\author[1,2]{Philipp Dönges}
\author[1,2*]{Seba Contreras}

\affil[1]{Max Planck Institute for Dynamics and Self-Organization, G\"ottingen, Germany.}
\affil[2]{Institute for the Dynamics of Complex Systems, University of G\"ottingen, G\"ottingen, Germany.}
\affil[ ]{* Corresponding Author: Seba Contreras (seba.contreras@ds.mpg.de)}
\affil[ ]{{$\dagger$} These authors contributed equally}

\date{}

\renewcommand{\headeright}{ }
\renewcommand{\undertitle}{ }

\begin{document}
\maketitle
\vspace*{-1cm}
\begin{abstract}
HIV pre-exposure prophylaxis (PrEP) drastically reduces the risk of HIV infection if taken as prescribed, providing almost perfect protection even during unprotected sexual intercourse. Although this has been transformative in reducing new HIV infections among high-risk populations, it has also been linked to an increase in risk practices—a phenomenon known as risk compensation—thereby favoring the spread of other sexually transmitted infections (STIs) deemed less severe. In this paper, we study a minimal compartmental model describing the effect of risk awareness and risk compensation due to PrEP on the spread of other STIs among a high-infection-risk group of men who have sex with men (MSM). The model integrates three key elements of risk-mediated behavior and PrEP programs: (i) HIV risk awareness drives self-protective behaviors (such as condom use and voluntary STI screening); (ii) individuals on PrEP are subject to risk compensation, but (iii) are required to screen for asymptomatic STIs frequently. We derived the basic reproduction number of the system, $R_0$, and found a transcritical bifurcation at $R_0=1$, where the disease-free equilibrium becomes unstable and an endemic equilibrium emerges. This endemic equilibrium is asymptotically stable wherever it exists. We identified critical thresholds in behavioral and policy parameters that separate these regimes and analyzed typical values for plausible parameter choices. Beyond the specific epidemiological context, the model serves as a general framework for studying nonlinear interactions between behavioral adaptation, preventive interventions, and disease dynamics, providing insights into how feedback mechanisms can lead to non-trivial responses in epidemic systems. Finally, our model can be easily extended to study the effect of interventions and risk compensation in other STIs.
%
\end{abstract}
\clearpage
 
 \section{Introduction}

HIV Pre-Exposure Prophylaxis (PrEP) has transformed HIV control. By offering strong protection against HIV infection when administered daily, it reduces the risk of infection following sexual intercourse by up to 99\%\cite{anderson2012emtricitabine,marcus2019risk}. While its widespread adoption in high-income settings has substantially reduced the heightened HIV burden among risk populations, such as men who have sex with men (MSM) \cite{anderson2012emtricitabine, mccormack2016pre, rozhnova2018elimination, murchu2022oral}, PrEP uptake has been linked with shifts in sexual behavior, e.g. reduced condom use, in a phenomenon known as risk compensation \cite{ayerdi2021low, milam2019sexual, storholm2017risk, golub2010preexposure}. This means that PrEP users are more likely to engage in riskier sexual behavior, driven by a lower perceived risk of acquiring HIV and a general perception that, because other STIs are curable, they are less severe \cite{holt2010gay}. Such behavior can facilitate the spread of STIs such as chlamydia and gonorrhea, which are characterized by high proportions of asymptomatic infections and therefore spread largely undetected \cite{hoornenborg2018change, holt2018community, quaife2020risk, van2020diverging}. To counteract this effect, many PrEP programs mandate or recommend regular STI screening \cite{spinner2019summary}: first, screening enables the detection, treatment, and counseling of positive cases that would have been otherwise undetected, reducing the force of infection \cite{contreras2021challenges}, and second, it prompts behavioral changes that may reduce the spreading rate of such infections \cite{wagner2025societal,donges2022interplay}. Therefore, testing provides a powerful countervailing mechanism to risk compensation. Together, behavioral adaptation, evolving public health interventions, and interactions between different diseases illustrate the complex and interdependent dynamics that shape STI transmission---dynamics that are not yet fully understood.

While empirical work has illuminated aspects of testing policies and risk compensation, observational data alone can make it difficult to distinguish true epidemiological trends from surveillance artifacts \cite{testing_paradox}. Therefore, mathematical models incorporating behavioral components, testing, and their interactions with STI spread are needed to untangle these effects. Despite their importance, only a few studies have examined these entangled processes in detail \cite{walker2022assessing, jenness2017incidence, pharaon_impact_2020, escobar_agent-based_2015}. To meaningfully interpret such models, it is essential to understand the underlying system dynamics and stability structure. In particular, models that couple testing, risk compensation, and STI transmission can generate complex nonlinear feedbacks whose qualitative behavior remains poorly understood. While several mathematical studies have explored bifurcations and multistability in other epidemiological systems \cite{abiodun2022mathematical, saha2021global, jing2024backward, wang2022bifurcations, saha2025dynamic, pradeesh2025bistability, corcoran2021low}, similar analyses are largely absent for models that explicitly incorporate behavioral and surveillance mechanisms, as e.g., \cite{testing_paradox}. 

In this work, we analyze a minimal mechanistic compartmental model that elucidates the coupled transmission dynamics of HIV and curable STIs among MSM. The model integrates three core feedbacks: (i) risk-driven modification of protective behavior and voluntary asymptomatic testing; (ii) PrEP-induced reduction in protective behavior; and (iii) mandatory regular asymptomatic STI screening associated with PrEP programs. The manuscript is organized as follows. Section~2 introduces the modeling assumptions and governing differential equations. Section~3 proves the existence and positivity of solutions, as well as the numerical stability of the model. Section~4 presents fixed points, linear stability, and bifurcations in the model as a function of the determined basic reproduction number. Section~5 presents a numerical exploration of the phase space of combinations between PrEP uptake and risk awareness, as well as the determination of critical parameters that determine stability. Section~6 presents the discussions and outlook. This analysis provides practical insights for public health while, from a theoretical standpoint, demonstrating how coupled behavioral and policy feedbacks interact to fundamentally reshape the system's stability landscape and generate non-monotonic phenomena.

\newpage

\section{Mathematical model}

We formalize the concepts of risk-related behavioral adaptation and STI screening in a compartmental model of concurrent HIV and STI transmission among MSM, using chlamydia as a representative example of bacterial STIs. Given that the spreading dynamics of HIV and chlamydia have different characteristic times (e.g., as per their generation intervals and spreading rates, we resort to the method of timescale separation to simplify our analysis: As the focus is on the dynamics of chlamydia, HIV incidence and awareness do not vary substantially and are assumed to be constant. We study a population of high-infection-risk MSM, in the sense that the well-mixing assumption required for ODE models is appropriate (see, e.g., \cite{rozhnova2018elimination,testing_paradox} for details on the appropriateness of this assumption). The model is structured around two key parameters: the level of PrEP uptake, $P$, and the perceived HIV infection risk, $H$. These quantities shape STI transmission dynamics through three feedback mechanisms that capture how individuals adapt their behavior in response to risk perception and PrEP uptake:

\paragraph{Risk-driven modification of behavior}
For individuals not on PrEP, two forms of behavioral adaptation play a central role. First, they adhere to protective behavior based on their perceived HIV infection risk, $m(H)$, which effectively lowers the STI transmission rate (Eq.~\eqref{eq:forceofinfection}). Second, they undergo voluntary asymptomatic STI screening at a rate $\lambda_H(H)$ (Eq.~\eqref{eq:feedback_5}). Both $m(H)$ and $\lambda_H(H)$ are assumed to increase monotonically with the perceived risk $H$ \cite{wagner2025societal,donges2022interplay}.

\paragraph{Reduced protective behavior of PrEP users}
We assume that PrEP users assimilate just a part of the risk they can measure, an effect captured by the risk assimilation $\xi$, which scales $m(H)$ (Eq.~\eqref{eq:forceofinfection}). Here, $\xi=1$ means PrEP users protect themselves like non-PrEP users, while $\xi=0$ reflects no risk assimilation, i.e., full risk compensation. Lower $\xi$ (greater risk compensation) increases transmission and enlarges the infectious pool, whereas higher $\xi$ (more protective behavior) reduces it.

\paragraph{Mandatory screening for PrEP users}
Beyond behavioral adaptations, PrEP use also entails a policy-driven component: mandatory regular STI screening at a rate $\lambda_P$ as required by PrEP programs\,\cite{spinner2019summary}. The parameter $\lambda_P$ represents the frequency of mandatory STI testing among PrEP users per year. Its effects are more complex than those of $\xi$. Mandatory testing identifies hidden infections and moves individuals from the infectious to the recovered compartment, which should reduce the reproduction number. However, voluntary testing complicates this: non-PrEP users tend to test based on perceived risk, whereas PrEP users rely on mandatory tests and rarely seek additional testing. If voluntary testing rates exceed mandatory rates, higher PrEP uptake will reduce overall testing, potentially enlarging the infectious pool and thereby indirectly undermining STI control.

\begin{figure}[!h]
    \centering    \includegraphics[width=100mm]{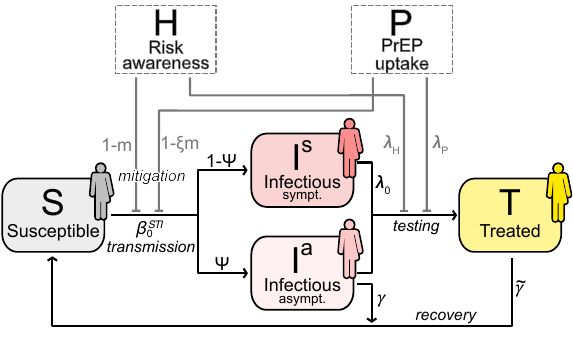}
    \caption{%
        \textbf{Minimal susceptible-infectious-treated-susceptible model for STI transmission among a high-infection-risk group of MSM, shaped by HIV risk awareness and PrEP uptake.} The model distinguishes between symptomatic and asymptomatic infections and incorporates three feedback mechanisms: i) increased awareness of HIV infection risk leads to higher condom use (mitigation $m$), ii) PrEP users reduce condom use (risk compensation, $\xi$), and iii) asymptomatic screening due to risk perception ($\lambda_H$, voluntary) and PrEP uptake ($\lambda_P$, mandatory). Not shown: an external influx of infections $\Sigma$, allocated between \Iasti and \Issti in proportion to the natural asymptomatic fraction ($\psi$) and its complement.
        }
    \label{fig:minimal_model}
\end{figure}

This differentiation of behavioral responses and screening frequency between PrEP users and non-users generates multiple feedback loops, both reinforcing and mitigating, that shape STI transmission and case detection across the population. A flow diagram of the model is shown in Fig.~\ref{fig:minimal_model}.

The model further distinguishes between asymptomatic and symptomatic infections, since these states affect both transmission and detection. Accordingly, the infectious compartment is split into $I^a$ (asymptomatic) and $I^s$ (symptomatic) individuals (Eqs.~\eqref{eq:dIasti_simple} and~\eqref{eq:dIssti_simple}, see Tab.~\ref{tab:variables}). Upon infection, individuals develop asymptomatic infection with probability $\psi$. Asymptomatic cases are primarily detected through screening ($\lambda_H(H)$ or $\lambda_P$) (Eq.~\eqref{eq:feedback_3}), whereas symptomatic cases are additionally detected through symptom-driven testing behavior at baseline rate $\lambda_0$ (Eq.~\eqref{eq:feedback_4}). All detected cases receive effective treatment, which removes them from the infectious pool and confers a temporary period of immunity, represented by compartment $T$ (Eq.~\eqref{eq:dTsti_simple}).

Finally, standard demographic processes are represented by a recruitment rate ($\Phi$) and an exit rate ($\mu$) from the sexually active population, which are analogous to birth and death rates in traditional models. To account for external introductions, we incorporate an influx $\Sigma$ of infectious individuals. An overview of all model parameters and default values can be found in table~\ref{tab:parameters}.

\begin{align}
\frac{d \Ssti}{dt}    & =  -\Lambda\Ssti+ \gamma \Iasti + \tilde{\gamma} \Tsti + \Phi  - \mu \Ssti - \Sigma  & : \quad& \text{susceptible,}\label{eq:dSsti_simple}\\
\frac{d \Iasti}{dt}    & =  \psi \cdot \Lambda\Ssti - \gamma \Iasti - \lambda_a(P,H)\Iasti - \mu \Iasti + \psi\Sigma   & : \quad& \text{asymptomatic infection,} \label{eq:dIasti_simple}\\
\frac{d \Issti}{dt}    & =  (1-\psi) \cdot \Lambda\Ssti - \lambda_s(P,H)\Issti - \mu \Issti+ (1-\psi)\Sigma& : \quad& \text{symptomatic infection,}\label{eq:dIssti_simple}\\
\frac{d \Tsti}{dt}    & =  \lambda_a(P,H) \Iasti + \lambda_s(P,H)\Issti- \tilde{\gamma} \Tsti - \mu \Tsti  & : \quad& \text{treated/recovered,}  \label{eq:dTsti_simple}
\end{align}

\begin{align}
        \Lambda & = \beta_{0}^{\textrm{STI}}\left((1 - m(H))(1 - P) + (1 - \xi \cdot m(H)) \cdot P\right) (\Iasti+\Issti) & : \quad& \text{force of infection,} \label{eq:forceofinfection}\\
    \lambda_{a}(P,H) & = \lambda_H(H) \cdot (1-\PrEP) + \lambda_P \cdot \PrEP & : \quad& \text{asymptomatic testing rate,} \label{eq:feedback_3}\\
\lambda_{s}(P,H) & = \lambda_0+\lambda_{a}(P,H),& : \quad& \text{symptomatic testing rate,} \label{eq:feedback_4}\\
\lambda_H(H) & =  k_\text{or}  \cdot \beta^{\text{HIV}}_0 (1-m(H))  H& : \quad& \text{risk-related testing rate,} \label{eq:feedback_5}\\
m(H) & =  m_{\textrm{min}} + (m_{\textrm{max}}-m_{\textrm{min}})\left(1-\exp\left(\frac{-H}{H_{\textrm{max}}}\right)\right)& : \quad& \text{mitigation function.} \label{eq:mitigation}
\end{align}

\begin{table*}[ht!]\caption{\textbf{Model variables.} All variables represent fractions of the population unless stated otherwise.}
\label{tab:variables}
\centering
\begin{tabular}{lp{10cm}}\toprule
\makecell[l]{Variable} & Definition  \\\midrule
$\Ssti$ & Fraction of the population susceptible to curable STIs  \\
$\Iasti$ & Fraction of the asymptomatic population infectious with curable STIs  \\
$\Issti$ & Fraction of the symptomatic population infectious with curable STIs  \\
$\Tsti$ & Fraction of the population treated against curable STIs  \\
$H$ & Fraction of the population that is risk aware  \\
$P$ & Fraction of the population that is on HIV PrEP  \\
\bottomrule
\end{tabular}%
\end{table*}

\begin{table*}[ht!]\caption{\textbf{Model parameters.}}
\label{tab:parameters}
\begin{tabular}{lp{8cm}lll}\toprule
Parameter & Definition   & Default value & Units & Source \\\midrule
$\beta_{0}^{\textrm{HIV}}$ & Base spreading rate of HIV within risk group & $0.6341$ & \SI{}{yr^{-1}}& \cite{rozhnova2018elimination}\\
$\beta_{0}^{\textrm{STI}}$ & Base spreading rate of chlamydia within risk group &  $0.008$& \SI{}{day^{-1}} & \cite{montes2022evaluating}\\
$\gamma$ & Chlamydia recovery rate (natural) & $1/1.32$ & \SI{}{yr^{-1}} & \cite{qu2021effect}\\
$\tilde{\gamma}$ & Chlamydia treatment-mediated recovery rate & 1/7 & \SI{}{day^{-1}} & \cite{qu2021effect}\\ 
$\lambda_0$ & Self-reporting rate for symptomatic STI infection & $1/l_{\textrm{STI}}$ & \SI{}{day^{-1}} & Calculated \\
$\lambda_a(P,H)$ & Testing rate (asymptomatic) for STI & Eq.~\eqref{eq:feedback_3}& \SI{}{day^{-1}} & Calculated\\
$\lambda_s(P,H)$ & Testing rate (symptomatic) for STI & Eq.~\eqref{eq:feedback_4}& \SI{}{day^{-1}} & Calculated\\
$\lambda_H (H)$ & Risk-related testing rate for STI or self-reporting & Eq.~\eqref{eq:feedback_5}& \SI{}{day^{-1}} & Calculated\\
$\lambda_P$ & PrEP-related screening rate for STI & [0,4] & \SI{}{yr^{-1}} & \cite{world2021consolidated,hevey2018prep}, scanned \\
$\mu$ & Exit rate from sexually active population & 1/45 & \SI{}{yr^{-1}} &\cite{rozhnova2018elimination} \\
$\xi$ & Risk assimilation & [0,1] & \si{-} & Scanned\\
$\Sigma$ & Total influx people infected with STI & $0.0$ & \SI{}{yr^{-1}} & Assumed\\
$\Phi$ & Recruitment rate to sexually active population & 1/45 & \SI{}{yr^{-1}} &\cite{rozhnova2018elimination} \\
$\psi$ & Fraction of asymptomatic chlamydia infection & $0.85$ & - & \cite{lamontagne2003chlamydia,korenromp2002proportion} \\
$H_{\textrm{max}}$ & Characteristic reaction awareness & $0.2$  & \SI{}{-} & Assumed\\
$k_\text{or}$ & Odds ratio of perceiving risk if one is at risk & 50 & - & \cite{kesler2016actual}\\
$l_{\textrm{STI}}$ & Incubation period for chlamydia infection in men & 14 & \SI{}{day} & \cite{korenromp2002proportion} \\
$m $ & Mitigation, self-regulation of contagious contacts & Eq.~\eqref{eq:mitigation} & - & Calculated\\
$m_{\textrm{min}}$ & Minimum mitigation & 0  & - & Assumed \\
$m_{\textrm{max}}$ & Maximum mitigation & 1  & - & Assumed \\
\bottomrule
\end{tabular}%
\end{table*}

\clearpage
\section{Basic properties of the model: positivity and boundedness}
All parameters are assumed to be nonnegative. Furthermore, $\psi, \xi, H, P \in[0,1]$. This ensures that  $\Lambda\ge 0$, $\lambda_a\ge 0$, and $\lambda_s\ge 0$ (Eqs.~
\eqref{eq:forceofinfection}--\eqref{eq:feedback_5}).

We focus on the model dynamics in the absence of external infection influx ($\Sigma=0$). A parallel analysis for scenarios with a non-zero influx ($\Sigma>0$) is provided in the Supplementary Material \ref{ssec:supplementary_methods}.
In case of a non-zero external influx, we assume it to be smaller than the recruitment rate
\begin{equation}
    \Sigma \leq \Phi\,\mathrm{.}
\label{eq:influx_condition}
\end{equation}

\medskip
\noindent\textbf{Theorem 1 (Positivity of solutions).}
\textit{Let $(\Ssti(0),\Iasti(0),\Issti(0),\Tsti(0))\in\mathbb{R}^4_{\ge 0}$. Then the solution $(\Ssti(t),\Iasti(t),\Issti(t),\Tsti(t))$ of Eqs.~\eqref{eq:dSsti_simple}--\eqref{eq:dTsti_simple} remains in $\mathbb{R}^4_{\ge 0}$ for all $t\ge 0$.}

\medskip
\noindent\textit{Proof.}
The vector field is locally Lipschitz continuous on $\mathbb{R}^4_{\ge 0}$ because all functions of parameters and $H,P$ are smooth, and all coefficients are bounded on compact sets; hence, a unique local solution exists and depends continuously on the initial value.

To show forward invariance of the nonnegative orthant, we verify the non-negativity barrier conditions compartment-wise.

(i) At $\Ssti=0$ with $\Iasti,\Issti,\Tsti\ge 0$,
\[
\frac{d \Ssti}{dt}\Big|_{\Ssti=0} \;=\; \gamma \Iasti + \tilde{\gamma} \Tsti + \Phi - \Sigma.
\]
Provided $\Sigma(0, \Iasti, \Issti, \Tsti) \leq \Phi+\gamma \Iasti+\tilde{\gamma\Tsti}$ at \Ssti=0, which follows that \\
\[
\frac{d \Ssti}{dt}\Big|_{\Ssti=0} \; \ge \; 0.
\]
since $\gamma,\tilde{\gamma},\Phi,\Sigma\ge 0$. Thus, $\Ssti$ cannot become negative.

(ii) At $\Iasti=0$ with $\Ssti,\Issti,\Tsti\ge 0$,
\[
\frac{d \Iasti}{dt}\Big|_{\Iasti=0} \;=\; \psi\,\Lambda \Ssti + \psi\,\Sigma \;\ge\; 0,
\]
since $\psi\in(0,1)$, $\Lambda\ge 0$, $S\ge 0$, and $\Sigma\ge 0$. Hence, $\Iasti$ cannot become negative.

(iii) At $\Issti=0$ with $\Ssti,\Iasti,\Tsti\ge 0$,
\[
\frac{d \Issti}{dt}\Big|_{\Issti=0} \;=\; (1-\psi)\,\Lambda \Ssti + (1-\psi)\,\Sigma \;\ge\; 0,
\]
since $1-\psi>0$, $\Lambda\ge 0$, $S\ge 0$, and $\Sigma\ge 0$. Thus, $\Issti$ cannot become negative.

(iv) At $\Tsti=0$ with $\Ssti,\Iasti,\Issti\ge 0$,
\[
\frac{d \Tsti}{dt}\Big|_{\Tsti=0} \;=\; \lambda_a \Iasti + \lambda_s \Issti \;\ge\; 0,
\]
since $\lambda_a,\lambda_s\ge 0$. Therefore, $\Tsti$ cannot become negative.

By Nagumo’s theorem \cite{nagumo1942lage}, or standard barrier arguments, the nonnegative orthant $\mathbb{R}^4_{\ge 0}$ is forward invariant, proving positivity. \qed

\noindent\textbf{Theorem 2 (Boundedness of solutions).}
\textit{
Assume  $\Phi=\mu \ge 0$. Given non-negative initial conditions, the total population $N(t) = \Ssti(t) + \Iasti(t) + \Issti(t) + \Tsti(t)$ satisfies $0\le N(t)\le \max\{1,N(0)\}$ for all $t\ge 0$.
Consequently,
\begin{align*}
0\le \Ssti(t),\,\Iasti(t),\,\Issti(t),\,\Tsti(t)\le N(t)\le \max\{1,N(0)\}\quad\text{for all }t\ge 0\,\mathrm{.} 
\end{align*}}

\medskip
\noindent\textit{Proof.}
Summing the four equations \eqref{eq:dSsti_simple}--\eqref{eq:dTsti_simple} gives
\begin{align}
\dot N=\Phi-\mu(\Ssti+\Iasti+\Issti+\Tsti)=\Phi-\mu N.
\label{ndot}
\end{align}
From Eq.~\eqref{mu_equal_phi}, we know $\Phi=\mu$, which reduces Eq.~\eqref{ndot} to
\begin{align}
\dot N=\mu(1-N).
\label{eq:linode}
\end{align}
For $\mu\ge 0$, Eq.~\eqref{eq:linode} is a linear first-order ODE, which implies that $N(t)$ remains nonnegative and $N(t)\le \max\{1,N(0)\}$. 
By Theorem~1 the components remain nonnegative, and since each component is $\le N(t)$ we obtain
\begin{align*}
0\le \Ssti(t),\,\Iasti(t),\,\Issti(t),\,\Tsti(t)\le N(t)\le\max\{1,N(0)\},
\quad\text{for all }t\ge0.
\end{align*}
This proves the stated bounds and completes the proof. \qed

\medskip
\noindent\textbf{Corollary (Invariant hyperplane and asymptotics for $\mu>0$).}\label{cor:N1}
\textit{
Assume $\Phi=\mu > 0$. Then the hyperplane $\{N=1\}$ is positively invariant and globally attracting.
Specifically, if $N(0)=1$ then $N(t)\equiv1$, and for any $N(0) \ge 0$, $N(t)\to1$ exponentially as $t\to\infty$.
Moreover, for $N(0)=1$, each compartment is bounded by $1$:
\begin{align*}
0\le \Ssti(t),\,\Iasti(t),\,\Issti(t),\,\Tsti(t)\le 1
\quad\text{for all }t\ge 0 \,\mathrm{.}
\end{align*}}

\medskip
\noindent\textit{Proof.}
Under the hypothesis $\Phi=\mu > 0$, the equation for $N(t)$ from the proof of Theorem 2 is $\dot N=\mu(1-N)$. Because $\mu > 0$, this linear ODE has the unique explicit solution:
\[
N(t)=1+(N(0)-1)e^{-\mu t}.
\]
If $N(0)=1$, then $N(t)\equiv1$ for all $t \ge 0$, which proves the positive invariance of the hyperplane $\{N=1\}$.
If $N(0) \ne 1$, the solution $N(t)$ approaches $1$ exponentially at rate $\mu$, showing the hyperplane is globally attracting.
The final statement of the bounds $0 \le \Ssti, \Iasti, \Issti, \Tsti \le 1$ follows from $N(t) \equiv 1$ and Theorem 1 (positivity). \qed

\medskip

We consider the model equations \eqref{eq:dSsti_simple}--\eqref{eq:dTsti_simple}
under the assumption of a constant total population, which is further normalized to one
\begin{equation}
    \Ssti + \Iasti + \Issti + \Tsti = N = 1.
    \label{eq:demographics}
\end{equation}
This condition implies demographic equilibrium (i.e., the sum of all derivatives is zero), requiring the recruitment rate $\Phi$ to be equal to the exit rate $\mu$
\begin{equation}
\Phi = \mu \left(\Ssti + \Iasti + \Issti + \Tsti\right) = \mu \cdot 1 = \mu.
\label{mu_equal_phi}
\end{equation}
This normalization allows for a reduction of the system from four to three differential equations, as $\Tsti$ can be expressed as an algebraic function of the other state variables
\begin{equation} \Tsti = 1 - \left(\Ssti + \Iasti + \Issti\right). \end{equation}

\noindent\textbf{Remark.} On the invariant hyperplane $\{N=1\}$,\; $\Tsti=1-(\Ssti+\Iasti+\Issti)$, so the reduced $(\Ssti,\Iasti,\Issti)$-system inherits positivity by the same boundary arguments on $\Ssti,\Iasti,\Issti$, and boundedness follows from $0\le \Ssti+\Iasti+\Issti\le 1$ implying $0\le T\le 1$. \hfill\(\square\)

\section{Fixed points and stability analysis}

\subsection*{Calculation of fixed points}

The fixed points ($S^*$, $I_a^*$, $I_s^*$) are determined by setting the time derivatives to zero: $\frac{d}{dt}\left(S^*, I_a^*, I_s^*\right)=(0,0,0)$. In the absence of external influx of infected individuals (i.e., $\Sigma=0$), the steady-state equations are:
\begin{align}
0 &= -\kappa \left(I_a^*+I_s^*\right) S^* + \gamma I_a^* + \tilde{\gamma} \left(1 - (S^* + I_a^* + I_s^*)\right) + \mu - \mu S^* \, ,\label{fpeq:1_main_sigma0} \\
0 &= \psi \kappa\left(I_a^*+I_s^*\right) S^* - (\gamma + \lambda_a + \mu) I_a^* \, , \label{fpeq:2_main_sigma0} \\
0 &= (1 - \psi) \kappa\left(I_a^*+I_s^*\right) S^* - (\lambda_s + \mu) I_s^* \label{fpeq:3_main_sigma0} \, ,
\end{align}
where
\begin{align}
    \kappa = \beta_0^{\mathrm{STI}}\left((1 - m(H))(1 - P) + (1 - \xi \cdot m(H)) \cdot P\right) = \beta_0^{\mathrm{STI}}\left(1 - m(H)\cdot( 1 - P\cdot(1 - \xi) )\right).
\label{eq:kappa}
\end{align}
From Eq.~\eqref{fpeq:3_main_sigma0}, assuming an endemic state where $I_a^* + I_s^* \neq 0$ and $1-\psi \neq 0$, $S^*$ can be expressed as:
\begin{equation}
S^* = \frac{ (\lambda_s + \mu) I_s^* }{ (1 - \psi) \kappa (I_a^* + I_s^*) }\, .\label{fpeq:S_main_sigma0}
\end{equation}
The relationship between $ I_s^* $ and $ I_a^* $ follows from Eqs.~\eqref{fpeq:2_main_sigma0} and \eqref{fpeq:3_main_sigma0}:
\begin{equation}
I_s^* = \frac{ (\gamma + \lambda_a + \mu) (1 - \psi) }{ \psi (\lambda_s + \mu) } I_a^*  \overset{\underset{\mathrm{def}}{}}{=} C_2  I_a^* \, .\label{fpeq:Is_Ia_main_sigma0}
\end{equation}
The constant $C_2$ is positive, given that all model parameters are positive and $0 < \psi < 1$.\\
Using Eqs.~\eqref{fpeq:1_main_sigma0}, \eqref{fpeq:S_main_sigma0}, and \eqref{fpeq:Is_Ia_main_sigma0}, we obtain an equation for $ I_a^* $:
\begin{equation}
    a(I_a^*)^2 + bI_a^* = 0\, ,\label{eq:quadratic_for_Ia_main_sigma0}
\end{equation}
where the coefficients are:
\begin{align}
a &= \kappa (1 + C_2) \left[ - (\lambda_s + \mu) C_2 + (\gamma - \tilde{\gamma} (1 + C_2))(1 - \psi) \right]\, , \label{fpeq:a_main_sigma0} \\
b &= (\tilde{\gamma} + \mu) \left[\kappa (1 - \psi) (1 + C_2) - (\lambda_s + \mu) C_2 \right] \, .\label{fpeq:b_main_sigma0} \\
\end{align}
Due to its quadratic term, Eq.~\eqref{eq:quadratic_for_Ia_main_sigma0} yields two solutions for $I_a^*$:
\begin{itemize}
    \item The zero solution for $I_a^*$ corresponds to the Disease-Free Equilibrium (DFE), which implies: 
        \begin{equation}
            (S^*, I_a^*, I_s^*) = (1, 0, 0).
        \end{equation}
    \item The non-zero solution for $I_a^*$ corresponds to the Endemic Equilibrium (EE). For $I_a^*$ to be positive and biologically meaningful, given that $a<0$ (which in the parameter regime we explore is true even for the case $C_2 = 0$), we require $b>0$. As discussed in the Bifurcation Analysis (sec.\ref{subsection:bifurcation-analysis}), $b>0$ if and only if $R_0 > 1$, as defined in Eq.~\ref{seq:R_0_main}. Therefore, the solution is:
    \begin{equation}
        (S^*, I_a^*, I_s^*) = \left(\frac{ (\lambda_s + \mu) C_2 }{ (1 - \psi) \kappa (1+C_2) }, -\frac{a}{b}, -C_2\frac{a}{b}\right)
    \end{equation}
\end{itemize}
Therefore, in the absence of influx, a DFE always exists, and an endemic equilibrium with $I_a^* > 0$ exists if and only if $R_0 > 1$.

\subsection*{Calculation of basic reproduction number $R_0$} \label{subsection:R_0}

The basic reproduction number, $R_0$, is derived using the next-generation matrix (NGM) method \cite{diekmann2010next}. This analysis focuses on new infections in the asymptomatic ($\Iasti$) and symptomatic ($\Issti$) compartments at the DFE, which for $\Sigma=0$ is $(\Ssti=1, \Iasti=0, \Issti=0)$. The linear approximation of the infection dynamics can be written as
\begin{equation}
        \frac{d}{dt}
    \begin{pmatrix}
        \Iasti \\
        \Issti
    \end{pmatrix}
    = \mathbf{F}(\Iasti, \Issti) - \mathbf{V}(\Iasti, \Issti), \label{seq:linearized_infection_subsystem_main}
\end{equation}
where $\mathbf{F}$ is the matrix of new infection rates and $\mathbf{V}$ is the matrix of transfer rates out of these compartments.\\
The matrices, evaluated at the DFE, are:
\begin{align}
    F &=
    \begin{pmatrix}
        \psi \kappa & \psi \kappa \\
        (1-\psi)\kappa & (1-\psi)\kappa
    \end{pmatrix},
    \label{seq:matrixF}
    \\
    V &=
    \begin{pmatrix}
        \Gamma_a & 0 \\
        0 & \Gamma_s
    \end{pmatrix},
    \label{seq:matrixV}
\end{align}
with 
\begin{align}
    \Gamma_a &= \gamma + \lambda_a + \mu &: \quad& \text{asymptomatic outflow rate}, \label{seq:asymptomatic_outflow} \\
    \Gamma_s &= \lambda_s + \mu &: \quad& \text{symptomatic outflow rate}. \label{seq:symptomatic_outflow}
\end{align} 

The inverse of $V$ is:
\begin{equation}
    V^{-1} =
    \begin{pmatrix}
        1/\Gamma_a & 0 \\
        0 & 1/\Gamma_s
    \end{pmatrix}.
\end{equation}
The next-generation matrix $ FV^{-1}$ is calculated as:
\begin{align}
    FV^{-1} &=
    \begin{pmatrix}
        \psi \kappa & \psi \kappa \\
        (1-\psi)\kappa & (1-\psi)\kappa
    \end{pmatrix}
    \begin{pmatrix}
        1/\Gamma_a & 0 \\
        0 & 1/\Gamma_s
    \end{pmatrix} \\
    &=
    \begin{pmatrix}
        \psi \kappa / \Gamma_a & \psi \kappa / \Gamma_s \\
        (1-\psi)\kappa / \Gamma_a & (1-\psi)\kappa / \Gamma_s
    \end{pmatrix}.
\end{align}
$R_0$ is the spectral radius $\rho(FV^{-1})$. The characteristic equation, $\det(FV^{-1} - \nu I) = 0$, is:
\begin{equation}
    \left(\frac{\psi \kappa}{\Gamma_a} - \nu\right) \left(\frac{(1-\psi)\kappa}{\Gamma_s} - \nu\right) - \left(\frac{\psi \kappa}{\Gamma_s}\right) \left(\frac{(1-\psi)\kappa}{\Gamma_a}\right) = 0,
\end{equation}
which simplifies to $\nu^2 - \nu \left(\frac{\psi \kappa}{\Gamma_a} + \frac{(1-\psi)\kappa}{\Gamma_s}\right) = 0$.\\
The non-zero eigenvalue is:
\begin{equation}
    R_0 = \frac{\psi \kappa}{\Gamma_a} + \frac{(1-\psi)\kappa}{\Gamma_s}, \label{seq:R_0_main}
\end{equation}
where the two terms reflect the expected number of secondary cases generated by asymptomatic and symptomatic individuals, respectively, in a wholly susceptible population.

This expression serves as a threshold: for $R_0 < 1$ the disease-free equilibrium is stable and the infection cannot invade, whereas $R_0 > 1$ allows for the possibility of disease persistence \cite{diekmann2010next}.

\subsection*{Linear Stability Analysis}

The local stability of the Disease-Free Equilibrium (DFE) and the Endemic Equilibrium (EE) for the system without influx ($\Sigma=0$) is determined by analyzing the eigenvalues of the Jacobian matrix $J$. This matrix is derived from the reduced system of differential equations for $(\Ssti, \Iasti, \Issti)$ and evaluated at the specific equilibrium point. The general form of the Jacobian is:
\begin{equation}
J = \begin{pmatrix}
-(\Lambda^*+\mu+\tilde{\gamma}) & \gamma-(\tilde{\gamma}+\kappa S^*) & -(\tilde{\gamma} + \kappa S^*) \\
\psi \Lambda^* & \kappa \psi S^* - \Gamma_a & \kappa\psi S^* \\
(1-\psi)\Lambda^* & \kappa (1-\psi) S^* & \kappa (1-\psi)S^* -\Gamma_s
\end{pmatrix},
\label{eq:fulljac_main_sigma0}
\end{equation}
where $\Lambda^* = \kappa(I_a^*+I_s^*)$ is the force of infection evaluated at the equilibrium $(S^*, I_a^*, I_s^*)$, and $\Gamma_a$ and  $\Gamma_s$ defined as before.

\subsubsection*{Stability of the Disease-Free Equilibrium (DFE)}
At the DFE, $(S^*=1, I_a^*=0, I_s^*=0)$, the force of infection $\Lambda^*$ is zero. Consequently, the Jacobian matrix simplifies to $J_{DFE}$:
\begin{equation}
J_{DFE} = \begin{pmatrix}
-(\mu+\tilde{\gamma}) & \gamma-(\tilde{\gamma}+\kappa) & -(\tilde{\gamma} + \kappa) \\
0 & \kappa \psi - \Gamma_a & \kappa\psi \\
0 & \kappa (1-\psi) & \kappa (1-\psi) -\Gamma_s
\end{pmatrix}.
\label{eq:jacDFE_main_sigma0}
\end{equation}
Due to the block triangular structure, one eigenvalue is readily identified as $-(\mu+\tilde{\gamma})$, which is always negative. The stability with respect to the introduction of infection is determined by the eigenvalues of the lower-right $2 \times 2$ submatrix, $M$:
\begin{equation}
M = \begin{pmatrix}
\kappa \psi - \Gamma_a & \kappa\psi \\
\kappa (1-\psi) & \kappa (1-\psi) -\Gamma_s
\end{pmatrix}.
\end{equation}
This submatrix $M$ is identical to $F-V$ (see Eq.~\eqref{seq:matrixF} and \eqref{seq:matrixV}). The DFE is locally asymptotically stable if all eigenvalues of the Jacobian matrix, evaluated at the DFE, have negative real parts. For a $2 \times 2$ system, this condition is equivalent to satisfying the Routh-Hurwitz stability criteria, which require $\mathrm{Tr}(M) < 0$ and $\mathrm{Det}(M) > 0$, where $\mathrm{Tr}(M) = \lambda_1 + \lambda_2$ and $\mathrm{Det}(M) = \lambda_1 \lambda_2$.

The equations for the trace and determinant can be written as: 
\begin{align} 
\label{eq:trace}
\mathrm{Tr}(M) &= \kappa\psi - \Gamma_a + \kappa(1-\psi) - \Gamma_s  \nonumber \\ &= \kappa - (\Gamma_a + \Gamma_s),\\
\label{eq:det}
    \mathrm{Det}(M) &= (\kappa\psi - \Gamma_a)(\kappa(1-\psi) - \Gamma_s) - \kappa^2\psi(1-\psi) \nonumber \\ 
    &= \Gamma_a\Gamma_s - \kappa\psi\Gamma_s - \kappa(1-\psi)\Gamma_a \nonumber \\
    &= \Gamma_a\Gamma_s (1-R_0).
\end{align}

Let us begin with the first condition, $\mathrm{Tr}(M) < 0$. Following Eq.~\eqref{eq:trace}, this is equivalent to $\kappa < \Gamma_a + \Gamma_s$. 

For this, we can rearrange Eq.~\eqref{seq:R_0_main} and use that we know $R_0<1$ for the disease-free equilibrium:

$$
\kappa < \frac{1}{\frac{\psi}{\Gamma_a} + \frac{(1-\psi)}{\Gamma_s}}.
$$

Given this, the condition $\kappa < \Gamma_a + \Gamma_s$ will hold if we show that $\frac{1}{\frac{\psi}{\Gamma_a} + \frac{(1-\psi)}{\Gamma_s}} < \Gamma_a + \Gamma_s$. Rearranging gives us

$$
 \frac{1}{\frac{\psi}{\Gamma_a} + \frac{(1-\psi)}{\Gamma_s}} < \Gamma_a + \Gamma_s \Leftrightarrow \frac{1}{\psi\Gamma_s+ (1-\psi)\Gamma_a} < \frac{1}{\Gamma_s} + \frac{1}{\Gamma_a}\quad .
$$

Noting that the function $f(x) = \frac{1}{x}$ is convex for $x>0$, we know that $f(\psi a + (1-\psi) b \leq \psi f(a) + (1-\psi) f(b)$, with $a,\,b\geq 0$. Applying this to $a=\Gamma_s$, $b=\Gamma_a$ we obtain

$$
\frac{1}{\psi\Gamma_s+ (1-\psi)\Gamma_a} \leq \frac{\psi}{\Gamma_s} + \frac{1-\psi}{\Gamma_a}<\frac{1}{\Gamma_s} + \frac{1}{\Gamma_a}.
$$
This proves that $\mathrm{Tr}(M) < 0$.

We will continue to prove the second Routh-Hurwitz condition, $\mathrm{Det}(M) > 0$. This is trivially given as we have $R_0<1$, which is the only requirement for Eq.~\eqref{eq:det} to be positive, given that $\Gamma_s, \Gamma_a>0$.

In conclusion, the DFE is locally asymptotically stable if $R_0 < 1$. On the contrary, if $R_0 > 1$, $\mathrm{Det}(M) < 0$, implying at least one positive eigenvalue, and the DFE becomes unstable.

\subsubsection*{Stability of the Endemic Equilibrium (EE)}
The existence of a unique, biologically feasible endemic equilibrium $(S_{EE}^*, I_{a,EE}^*, I_{s,EE}^*)$, characterized by $I_{a,EE}^* > 0$, has been established for the condition $R_0 > 1$ when $\Sigma=0$ (refer to Section~\ref{subsection:bifurcation-analysis}). To ascertain its local stability, we evaluate the Jacobian matrix $J_{EE}$ by substituting these endemic fixed-point coordinates into the general form given by Eq.~\eqref{eq:fulljac_main_sigma0}.

The stability of this equilibrium is governed by the eigenvalues $\nu$ of $J_{EE}$, which are the roots of the characteristic polynomial $\det(J_{EE} - \nu I) = 0$. For a $3 \times 3$ system, this yields a cubic equation:
\begin{equation}
\nu^3 + c_2 \nu^2 + c_1 \nu + c_0 = 0, \label{eq:char_poly_EE_main_revised}
\end{equation}
where the coefficients are standard functions of the elements of $J_{EE}$: $c_2 = -\mathrm{Tr}(J_{EE})$, $c_1$ is the sum of the principal minors of $J_{EE}$, and $c_0 = -\mathrm{Det}(J_{EE})$.

Local asymptotic stability of the endemic equilibrium requires that all roots of Eq.~\eqref{eq:char_poly_EE_main_revised} possess negative real parts. The Routh-Hurwitz stability criteria provide necessary and sufficient conditions for this:
\begin{align}
    c_2 > 0,\, c_1 > 0,\, c_0 > 0,\, c_1 c_2 - c_0 > 0 \nonumber
\end{align}
It is a well-established result for epidemiological models exhibiting a transcritical bifurcation at $R_0=1$ that the emergent endemic equilibrium gains stability as the disease-free equilibrium loses it. Accordingly, for this system, detailed analysis of the Routh-Hurwitz criteria confirms that all conditions are satisfied for the unique positive endemic equilibrium when $R_0 > 1$. Therefore, the endemic equilibrium is locally asymptotically stable whenever it exists.

\subsection*{Bifurcation Analysis}
\label{subsection:bifurcation-analysis}
The bifurcation behavior of the system in the absence of external influx ($\Sigma = 0$) is critically dependent on the basic reproduction number, $R_0$. As established previously, the DFE, $(\Ssti = 1, \Iasti = 0, \Issti = 0)$, is locally asymptotically stable if $R_0 < 1$ and becomes unstable if $R_0 > 1$.\\
The coefficient $b$, originally defined in Eq.~\eqref{fpeq:b_main_sigma0}, can be directly related to $R_0$. Through algebraic substitution of the definitions of $C_2$, $\Gamma_a$, $\Gamma_s$, and $R_0$, it can be written as:
\begin{equation}
    b = (\tilde{\gamma} + \mu) \frac{(1-\psi)\Gamma_a}{\psi} (R_0-1).
    \label{eq:b_vs_R0_main}
\end{equation}
The pre-factor $(\tilde{\gamma} + \mu) \frac{(1-\psi)\Gamma_a}{\psi}$ consists of positive model parameters (assuming $0 < \psi < 1$). Consequently, the sign of $b$ is solely determined by the sign of the term $(R_0-1)$:
\begin{itemize}
    \item If $R_0 < 1$, then $(R_0-1) < 0$, which implies $b < 0$. In this case, $I_a^* = -b/a < 0$, so no feasible positive endemic equilibrium exists.
    \item If $R_0 > 1$, then $(R_0-1) > 0$, which implies $b > 0$. In this case, $I_a^* = -b/a > 0$, indicating the existence of a unique, positive endemic equilibrium.
    \item If $R_0 = 1$, then $(R_0-1) = 0$, which implies $b = 0$. In this case, $I_a^* = 0$, and the DFE and the emerging endemic equilibrium coincide.
\end{itemize}
Thus, a unique, biologically feasible endemic equilibrium emerges and exists if and only if $R_0 > 1$.

Furthermore, at this endemic equilibrium (when $R_0 > 1$ and $\Sigma=0$), substituting the endemic fixed point values back into the system equations reveals that the susceptible fraction simplifies to:
\begin{equation}
    S_{EE}^* = \frac{1}{R_0}.
    \label{eq:SEE_is_1_over_R0_main}
\end{equation}
This change in stability of the DFE and the concurrent emergence of a stable endemic equilibrium as $R_0$ crosses the threshold value of 1 is characteristic of a transcritical bifurcation. The point $R_0=1$ is therefore the critical bifurcation point for this system.

\clearpage
\section{Results}
\subsection*{Stability as a trade-off between risk-adaptive behavior and PrEP screening frequency}

As shown in the previous section, the stability of the disease-free equilibrium depends on multiple factors, which can be grouped into two categories: risk-adaptive behavior (e.g., the behavioral response to risk perception $H$ or risk assimilation $\xi$) and structural factors, arising from population characteristics (e.g., PrEP uptake) or policy (e.g., PrEP screening frequency $\lambda_P$). These factors interact in nontrivial ways, as reflected in the functional shapes of the equilibria and the basic reproduction number. 
To illustrate these non-trivial dynamics, we visualize the basic reproduction number as a function of PrEP uptake ($P$) and risk awareness ($H$), for different combinations of $\xi$ (rows) and $\lambda_P$ (columns) (Fig.~\ref{fig:Equil_of_P_H_different_xi_and_lambdaP}). The regions of disease-free equilibrium (DFE, $R_0<1$) and endemic equilibrium (EE, $R_0>1$) are separated by a turquoise line, which corresponds to $R_0=1$ (Fig.~\ref{fig:Equil_of_P_H_different_xi_and_lambdaP}, a--i), and summarized in the stability plots in panels Fig.~\ref{fig:Equil_of_P_H_different_xi_and_lambdaP}j--l.

As expected, the area of the DFE expands with increasing $\xi$. This effect is especially pronounced for $\lambda_P=\SI{1}{year^{-1}}$ (Fig.~\ref{fig:Equil_of_P_H_different_xi_and_lambdaP}c,f,i), where high risk assimilation (i.e., no risk compensation among PrEP users) allows STIs to disappear once risk awareness exceeds approximately 0.08 (Fig.~\ref{fig:Equil_of_P_H_different_xi_and_lambdaP}i). In contrast, when risk assimilation is low, even high risk awareness cannot prevent STI persistence if PrEP uptake is high.

The role of the mandatory testing frequency $\lambda_P$ shows a different pattern. With high testing frequency, e.g. four times per year (Fig.~\ref{fig:Equil_of_P_H_different_xi_and_lambdaP}a,d,g), the endemic region is small, confined to combinations of both low PrEP uptake and low risk awareness. Increasing either variable leads to more testing (through either voluntary uptake or mandatory screening) and can drive elimination. With moderate testing (Fig.~\ref{fig:Equil_of_P_H_different_xi_and_lambdaP}b,e,h), the endemic region extends farther to the right as higher PrEP uptake produces only modest increases in testing, delaying elimination. For low testing frequencies ($\lambda_P=\SI{1}{year^{-1}}$, Fig.~\ref{fig:Equil_of_P_H_different_xi_and_lambdaP}c,f,i), the shape of the EE region changes markedly. The $R_0=1$ boundary shifts upward with increasing PrEP uptake, even sometimes bending backwards again (panel c).

The back-bending is because mitigation and risk-related testing both depend on risk awareness $H$, but in nonlinear ways. Mitigation increases with $H$ and gradually saturates at $H_\text{max}=0.2$, so that additional increases in awareness have diminishing effects at high $H$ (Eq.~\eqref{eq:mitigation}). Risk-related testing, in contrast, first rises with $H$ as more individuals recognize their exposure, but after reaching $H_\text{max}$, it declines again (Eq.~\eqref{eq:feedback_5}). This decline arises because testing is proportional to $(1 - m(H))H$: as mitigation $m(H)$ increases, fewer people engage in risky behavior and thus test less frequently.  

When PrEP uptake ($P$) is low, mitigation is strong enough to compensate for the eventual decline in testing, so increasing $H$ consistently lowers $R_0$. At intermediate levels of $P$, however, the $R_0 = 1$ contour develops a back-bending shape. In this regime, $R_0$ initially decreases with $H$ because both mitigation and testing increase, but as $H$ continues to rise, testing declines while mitigation saturates. Since PrEP users mitigate less than non-users, the resulting reduction in testing can no longer be offset, leading $R_0$ to rise again. Consequently, for a given $P$, two levels of $H$ can yield $R_0 = 1$. For high $P$, nearly everyone is on PrEP, so changes in $H$ have little effect. The back-bending disappears when PrEP users show little risk compensation (high $\xi$), meaning their mitigation remains comparable to that of non-users, or when PrEP-related testing is frequent enough to keep overall testing rates high even as risk-related testing declines. 

In summary, higher risk assimilation and PrEP-related testing frequency both consistently reduce the reproduction number, and therefore STI prevalence. Risk awareness, on the other hand, shows more complex patterns with rising awareness first reducing and then increasing prevalence again.

\begin{figure}
    \centering
    \includegraphics{Figures/Results_Fig_1.pdf}
    \caption{\textbf{a--i} Basic reproduction number as a function of risk awareness $H$ and PrEP uptake $P$ for different values of risk assimilation $\xi$ (rows) and mandatory testing frequencies for PrEP users $\lambda_P$ (columns). $R_0 < 1$ corresponds to the stable disease-free equilibrium, while $R_0 > 1$ corresponds to the endemic equilibrium. They are separated by $R_0=1$, which is highlighted in turquoise. \textbf{j--l} Comparison of the $R_0=1$ lines for the different testing frequencies. }
    \label{fig:Equil_of_P_H_different_xi_and_lambdaP}
\end{figure}

\clearpage
\subsection*{Critical $\lambda_P$ and $\xi$}

We investigate the parameter values that delimit the transition between DFE and EE. They will be referred to as the critical testing frequency and critical risk assimilation, denoted $\lambda_{P,\text{crit}}$ and $\xi_{\text{crit}}$, respectively. In other words, when $\lambda_P > \lambda_{P,\text{crit}}$ or $\xi > \xi_{\text{crit}}$, the system remains in the DFE ($R_0 < 1$), whereas for lower values it transitions to the EE ($R_0 > 1$). These critical thresholds are therefore key for identifying feasible and effective strategies for disease control and elimination.

We determine $\lambda_{P,\text{crit}}$ by fixing $\xi$ and varying $\lambda_P$, and analogously determine $\xi_{\text{crit}}$ by fixing $\lambda_P$ and varying $\xi$ (see Supplementary Material \ref{ssec:: supplementary_critical_parameters}). In Fig.~\ref{fig:Equil_of_P_H_different_xi_and_lambdaP}, this corresponds to fixing a row (for $\lambda_P$) or a column (for $\xi$) and then moving horizontally (increasing $\xi$) or vertically (increasing $\lambda_P$) to identify the point where the system shifts from EE to DFE. This point defines the critical value.

\begin{figure}[h]
    \centering
    \includegraphics{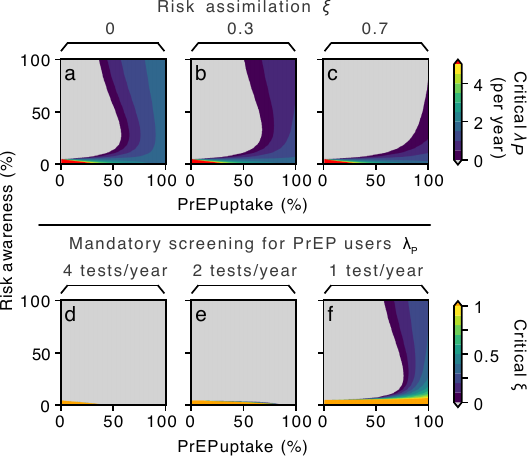}
    \caption{\textbf{Critical values of risk assimilation and testing frequency needed to go from endemic to disease-free equilibrium.} Panels \textbf{a–c} show the critical testing frequency $\lambda_{P,\text{crit}}$ and panels \textbf{d–f} the critical risk assimilation $\xi_\text{crit}$ based on risk awareness $H$ and PrEP uptake $P$. Colors represent levels of testing frequency (a–c) or risk assimilation (d–f). Grey indicates regions where the disease-free equilibrium always dominates, red marks unfeasibly high testing frequencies, and orange marks where the system is always in the endemic equilibrium.}
    \label{fig:critical}
\end{figure}

For $\lambda_{P,\text{crit}}$ (Fig.~\ref{fig:critical}a--c), we can distinguish two main regimes. In some parameter regions, the system remains in the DFE for all $\lambda_P$, regardless of how small it is (gray areas in Fig.~\ref{fig:critical}). In other regions, a finite critical value $\lambda_{P,\text{crit}}$ exists (colored areas). The color scale indicates the value of this critical testing frequency required to reach $R_0 = 1$. The spatial pattern of $\lambda_{P,\text{crit}}$ closely reflects the distribution of $R_0$ shown in Fig.~\ref{fig:Equil_of_P_H_different_xi_and_lambdaP}: in regions where $R_0$ was higher for a fixed low $\lambda_P$, a larger increase in testing is needed to reduce it to one, leading to higher $\lambda_{P,\text{crit}}$ values. Conversely, where $R_0$ was already close to one, only modest increases in testing suffice, yielding lower $\lambda_{P,\text{crit}}$. Red regions correspond to parameter combinations where the required testing frequency exceeds feasible levels (more than five tests per year). This occurs in regions where both PrEP uptake and risk awareness are low.

For $\xi_{\text{crit}}$ (Fig.~\ref{fig:critical}d--f), three regimes can be distinguished. As before, gray regions indicate parameter combinations that are always in the DFE, while colored regions mark areas where a finite $\xi_{\text{crit}}$ exists. In contrast to $\lambda_P$, which is only practically limited by feasibility rather than by definition, $\xi$ is inherently bounded between 0 and 1, which introduces a third regime (orange): parameter regions where even the maximum possible value $\xi = 1$ is insufficient to achieve $R_0 < 1$, and the system therefore remains in the EE for all $\xi$. As with $\lambda_{P,\text{crit}}$, the spatial patterns of $\xi_{\text{crit}}$ closely follow those of $R_0$ in Fig.~\ref{fig:Equil_of_P_H_different_xi_and_lambdaP}: where $R_0$ is higher for a low fixed $\xi$, stronger risk assimilation is required to reduce it to one, meaning that $\xi_{\text{crit}}$ is also high.

In summary, both higher testing rates and stronger risk assimilation promote the DFE, yet each faces distinct constraints---testing by practical feasibility and risk assimilation by behavioral limits. Mapping $\lambda_{P,\text{crit}}$ and $\xi_{\text{crit}}$ thus provides quantitative insight into the conditions under which elimination can be maintained and highlights which intervention dimension, testing or behavioral adaptation, offers the most leverage under different epidemiological contexts.

\clearpage
\subsection*{Relation between $\lambda_P$, $P$, and $H$}

\begin{figure}
    \centering
    \includegraphics{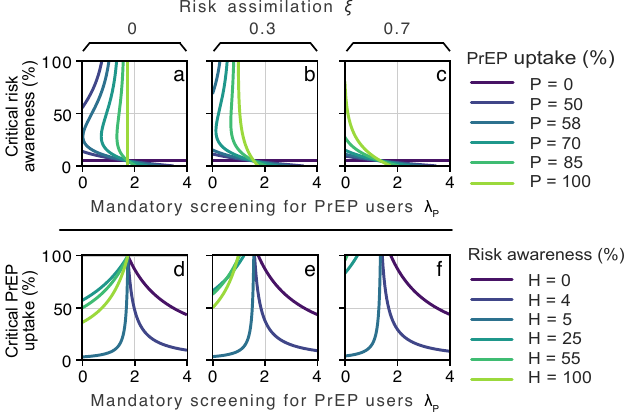}
    \caption{\textbf{Elimination thresholds marked by critical values as a function of mandatory testing frequency.} To the left and right of each line, the endemic or disease-free equilibrium dominates, respectively. Panels \textbf{a--c} show the critical risk awareness, $H_\text{crit}$, and panels \textbf{d--f} show the critical PrEP uptake, $P_\text{crit}$, as functions of the PrEP-related testing frequency, $\lambda_P$. Line colors represent levels of PrEP uptake (a--c) or risk awareness (d--f), and columns correspond to $\xi = 0$, $0.3$, and $0.7$.}
    \label{fig:critical_H_and_P}
\end{figure}

To better understand how the PrEP-related testing frequency $\lambda_P$ interacts with risk awareness $H$ and PrEP uptake $P$, we now examine how the critical level of these variables, defined as the level at which $R_0 = 1$, changes with $\lambda_P$. Fig.~\ref{fig:critical_H_and_P} illustrates these relationships: panels a--c show the critical risk awareness $H_{\text{crit}}$ as a function of $\lambda_P$ for different $P$, while panels d--f show the critical PrEP uptake $P_{\text{crit}}$ as a function of $\lambda_P$ for different $H$. The three columns correspond to increasing degrees of risk assimilation $\xi$.

Starting with $H_{\text{crit}}$ (Fig.~\ref{fig:critical_H_and_P}a--c), when PrEP uptake is maximal ($P=1$, light green line) and risk assimilation is zero ($\xi=0$), $R_0$ becomes independent of $H$, since all individuals are on PrEP and none mitigate. This means that $R_0=1$ corresponds to a vertical line, left and right of which the system remains entirely endemic or disease-free, respectively. As $P$ decreases (from green to blue), the curves shift leftward and bend, indicating that two $H$ values can yield $R_0=1$---consistent with the non-monotonic dependence of voluntary testing on risk awareness discussed above (Fig.~\ref{fig:Equil_of_P_H_different_xi_and_lambdaP}). At $P\approx 0.58$, the curve eventually touches the $y$-axis, meaning that even for $\lambda_P=\SI{0}{year^{-1}}$, elimination can be achieved if risk awareness is sufficiently strong. In the limit $P=0$, the model becomes independent of $\lambda_P$, resulting in a horizontal line, since PrEP-related testing no longer affects transmission. For risk awareness levels below and above this line, the system is in the endemic or disease-free state, respectively. Notably, all lines intersect at $\lambda_P\approx\SI{1.7}{year^{-1}}$---representing a critical $\lambda_P$ value at which the same $H_{\text{crit}}$ applies regardless of $P$. This marks the transition where, for lower $H$, $R_0$ decreases with $P$, while for higher $H$, $R_0$ increases with $P$ (Fig.~\ref{fig:Equil_of_P_H_different_xi_and_lambdaP}).

As $\xi$ increases (panels b--c), PrEP users begin to mitigate, so $R_0$ again depends on $H$ even when $P=1$. Higher $\xi$ shifts all curves leftward as mitigation increasingly compensates for less frequent testing suffices for elimination, and moves the intersection point to a smaller $\lambda_P$. Panels d--f show analogous trends for the critical PrEP uptake $P_{\text{crit}}$. For $\xi=0$, all lines meet at $P=1$ and the same $\lambda_P$ as the vertical line in panels a–c, reflecting independence from $H$ when risk assimilation is absent. With $\xi>0$, this convergence disappears, consistent with the disappearance of the vertical line in panels b--c.

For small $H$, $P_{\text{crit}}$ decreases with increasing $\lambda_P$, matching the region in Fig.~\ref{fig:Equil_of_P_H_different_xi_and_lambdaP} where $R_0$ falls with greater PrEP uptake. These curves begin only around $\lambda_P \approx \SI{2}{year^{-1}}$, since below this threshold the endemic equilibrium persists for all $P$, consistent with the results above. As $H$ rises, the curves exist only for smaller $\lambda_P$ and tend to rise, corresponding to regions where $R_0$ increases with $P$. Between these regimes, an intermediate $H$ produces a vertical line, corresponding to the horizontal $R_0=1$ contour in Fig.~\ref{fig:Equil_of_P_H_different_xi_and_lambdaP} and marking the same point where all lines meet in panels a--c.

Initially, a higher $H$ (going from blue to green) raises the required $P_{\text{crit}}$, but beyond a threshold (around $H\approx 0.25$), it declines again. This turning point corresponds to the bending behavior in panels a--c and Fig.~\ref{fig:Equil_of_P_H_different_xi_and_lambdaP}: as risk-related testing declines at high $H$, the endemic region expands toward smaller $P$, lowering the PrEP uptake required for $R_0 = 1$.

\clearpage
\section{Discussion}

We analyzed a compartmental model for the effect of risk awareness and risk compensation due to HIV PrEP on the spread of other STIs. This study complements our related work \cite{testing_paradox}, which employs a more complex formulation of the underlying model to investigate a "testing paradox" where the number of positive tests does not necessarily reflect true changes in underlying STI prevalence. This paper focuses on the mathematical foundations of the model and provides a detailed analysis of its mechanisms and modeling choices. Our analysis revealed a transcritical bifurcation: when $R_0 < 1$, only a disease-free equilibrium (DFE) exists and is stable, whereas when $R_0 > 1$, an endemic equilibrium (EE) emerges, rendering the DFE unstable. The exact point of bifurcation ($R_0=1$) depends primarily on four key parameters, which we analyze in detail below. 

The first key parameters are the HIV risk awareness ($H$) and risk assimilation ($\xi$). When $H$ is high, individuals not on PrEP mitigate more and screen more frequently, whereas this effect is modulated by $\xi$ for PrEP users. By construction, the effect of $H$ is more pronounced when $H\leq H_{\rm max} = 0.2$, as the functional shape of the mitigation function saturates afterwards. Realistic values for $H$ range between 0\% and 20\%, whereas $\xi$ is rather low, reflects consistent evidence from demographic and behavioral studies showing that a substantial proportion of high-risk MSM: (i) Do not perceive themselves to be at significant risk for HIV or other STIs; (2) Exhibit reduced condom use once on PrEP, despite remaining vulnerable to STIs  \cite{jiao2023mpox,khawcharoenporn2019hiv, balan2019low}.

The next parameters are the PrEP uptake ($P$) and the rate of PrEP-related asymptomatic screening for STIs ($\lambda_P)$. Changes in $P$ affect multiple aspects of the system. On the one hand, when risk assimilation is low, it decreases the fraction of the population that reacts to the risk perception $H$ and thus limits its stabilizing effect on the DFE. On the other hand, increasing $P$ modulates the effect of $\lambda_P$: if the mandatory screening frequency is high, the overall testing rate increases and it has a stabilizing effect on the DFE, which is lost otherwise. In settings with widespread PrEP uptake, such as in high-income European countries, PrEP screening frequency ranges from twice to four times a year \cite{clutterbuck20182016, centers2018preexposure, deutsche_aidshilfe_prepchecks}. However, asymptomatic screening for bacterial STIs, such as chlamydia and gonorrhea, is not always mandatory \cite{van2024belgian, white20252025}, with conflicting evidence on whether the reductions in incidence justify the overexposure to broad-spectrum antibiotics \cite{kenyon2023management, williams2023frequent}. It is important to clarify that within our framework, the concept of "treatment" extends beyond pharmacological intervention. It signifies an individual's exit from the infectious state, a transition driven by the comprehensive behavioral shifts that occur after diagnosis, supported by counseling and other non-pharmaceutical measures \cite{schnyder2025understanding,kettlitz2023association}. A natural extension would be to differentiate between the two pathways and assume imperfections in each.

From our model, we can draw two main actionable implications. First, education on sexual health plays a crucial role: both $H$ and $\xi$ can be modified through information campaigns and have a large impact on the stability of the DFE. Second, maintaining an adequate frequency of PrEP-related testing is essential. Across a wide range of parameter values, testing two or more times per year offers a favorable balance: it substantially shrinks the stability zone of the EE while remaining practical from an implementation standpoint. The situation could change with long-acting PrEP, as the administration frequency could interfere with the recommended screening frequency. Currently, administration frequency can vary from every two months for cabotegravir (currently approved in Europe \cite{ViiVHealthcare2022EMAcabotegravir}), to every six months for lenacapavir \cite{bekker2024twice}, or every month for islatravir \cite{hillier2021safety}, further adding complexity to the landscape \cite{wang2025oral}. However, given the simplicity of our model, we refrain from making a precise recommendation for the optimal testing frequency.

The assumptions in our model induce structural limitations that need to be considered. First, related to the population of study: the model aims to reflect the spread of STIs among a high-infection-risk group of MSM, so that the temporal variability in the network of sexual contacts, aggregated over the relevant timescales, can be effectively modeled under the well-mixing assumption (see e.g., \cite{rozhnova2016impact,rozhnova2018elimination,testing_paradox}). The assumption of homogeneous mixing within this group simplifies the system, allowing for a tractable analysis of its equilibrium and bifurcation structure. Second, we assume demographic equilibrium, i.e., that the recruitment and exit rates in the high-infection-risk group are in balance. However, the relationship between sexual behavior and life course is not linear \cite{khosropour2021changing, hess2017trends}, and it can even change on shorter timescales related to partnerships \cite{sandfort2013sexual,hoff2012relationship}. Finally, we assume that all positive STI tests result in prompt and effective treatment, although antimicrobial resistance is on the rise for STIs \cite{tien2020antimicrobial,de2022worryingly}, and that all individuals on PrEP take it as prescribed---reflecting both oral administration and long-acting PrEP. For Europe, although the intention of PrEP is high, there are still access barriers that need to be overcome to enhance current use and prepare for future LA-PrEP modalities \cite{wang2025oral}.

Future extensions could improve the model by incorporating additional layers of behavioral and demographic complexity. In particular, exploring the observability of the network of sexual contacts involves understanding that only a fraction of it can be observed, with varying degrees of confidence. Tools from subsampling theory will be crucial for understanding the dynamics within such networks \cite{wilting2018inferring,levina2017subsampling,levina2022tackling}. Coupling the current framework with data-informed parameter estimates or individual-based simulations may also improve its applicability to specific settings. Such extensions would enhance the model’s predictive utility and support the development of more targeted, evidence-based policies for STI and HIV prevention, helping to translate mathematical insights into actionable public health strategies.

\section*{Author Contributions}

Conceptualization: SC\\
Data curation: PM, LM, AM\\
Formal analysis: PM, AM\\
Investigation: PM, AM, LM, PD\\
Methodology: all\\
Project administration: SC\\
Software: PM, LM, AM\\
Supervision: SC\\	
Validation: all\\
Visualization: PM, LM, AM\\	
Writing - Original Draft: PM, LM\\	
Writing - Review \& Editing: all \\

\section*{Code availability}
All code to reproduce the analysis and figures shown in the manuscript, as well as in the Supplementary Material, is available online on Github \href{https://github.com/Priesemann-Group/FiND-STI}{https://github.com/Priesemann-Group/FiND-STI}. 

\section*{Acknowledgments} 

We thank Prof. Dr. Viola Priesemann for comments on this manuscript and our colleagues in the Priesemann Group for fruitful discussions. We thankfully acknowledge support from the Max-Planck Society and the Ministry of  Science and Culture of Lower Saxony, through funds from the program zukunft.niedersachsen of the Volkswagen Foundation, via the Niedersachsen-Profil-Professur, and via the 'CAIMed – Lower Saxony Center for Artificial Intelligence and Causal Methods in Medicine' project (grant no. ZN4257), of which the Priesemann group is part. ChatGPT and Grammarly AI were used for grammar checks in the main text, and GitHub Copilot served as a coding assistant. The authors assume full responsibility for the final content of the article.

\clearpage

\newpage
\renewcommand{\thefigure}{S\arabic{figure}}
\renewcommand{\figurename}{Supplementary~Figure}
\setcounter{figure}{0}
\renewcommand{\thetable}{S\arabic{table}}
\renewcommand{\tablename}{Supplementary~Table}

\setcounter{table}{0}
\renewcommand{\thesection}{S\arabic{section}}
\setcounter{section}{0}
\setcounter{page}{1}
\section{Supplementary Material} \label{SI}

\subsection{Critical parameters}\label{ssec:: supplementary_critical_parameters}

In the main text, we analyzed the critical values of the testing frequency, $\lambda_{P, \text{crit}}$, and risk assimilation, $\xi_{\text{crit}}$, at which a transcritical bifurcation occurs. This bifurcation marks the transition between the Disease-Free Equilibrium (DFE) and the Endemic Equilibrium (EE). In this section, we derive the implicit equation that defines this critical boundary.

The bifurcation from the DFE to the EE occurs precisely when the basic reproduction number is $R_0 = 1$. As established in the main text (Eq.~\eqref{eq:SEE_is_1_over_R0_main}), the susceptible population at the endemic equilibrium, $S^*_{EE}$, is related to $R_0$ by: 
\begin{equation}
    S^*_{EE} = \frac{1}{R_0}
\end{equation}
Therefore, at the critical boundary ($R_0=1$), the susceptible population is $S^*_{EE} = 1$.

Using the expression for $R_0$ from Eq.~\eqref{seq:R_0_main}, we can express $1/R_0$ (and thus $S^*_{EE}$) as:
\begin{equation}
    S^*_{EE} = \frac{1}{R_0} = \frac{ (\lambda_s + \mu) \Issti }{ (1 - \psi) \kappa (\Iasti + \Issti) }
\end{equation}
By setting $S^*_{EE} = 1$, we obtain the condition for the critical boundary:
\begin{equation}
    \frac{ (\lambda_s + \mu) \Issti }{ (1 - \psi) \kappa (\Iasti + \Issti) } = 1
\end{equation}

This expression can be rearranged to show the balance at the threshold:
\begin{equation}
    (\lambda_s + \mu) \Issti = (1 - \psi) \kappa (\Iasti + \Issti)
    \label{eq:crit_balance}
\end{equation}

To express this condition purely in terms of model parameters, we utilize the fixed-point relationship between the infected compartments from Eq.~\eqref{fpeq:Is_Ia_main_sigma0}. This relationship defines the ratio $C_2 = \Issti / \Iasti$, which is a function of $(P, H, \lambda_P)$.

Substituting $\Issti = C_2 \Iasti$ into Eq.~\eqref{eq:crit_balance} and dividing by $\Iasti$ (which is non-zero at the endemic equilibrium) yields:
\begin{align}
    (\lambda_s + \mu) (C_2 \Iasti) &= (1 - \psi) \kappa (\Iasti + C_2 \Iasti) \nonumber \\
    (\lambda_s + \mu) C_2 &= (1 - \psi) \kappa (1 + C_2)
\end{align}

This equation implicitly defines the critical boundary. We now write it explicitly in terms of the policy-dependent parameters at their critical values:
\begin{equation}
    (\lambda_s(P,H,\lambda_{P, \text{crit}})+\mu)C_2(P,H, \lambda_{P, \text{crit}}) = (1-\psi)\kappa(P, H, \xi_{\text{crit}})(1+C_2(P,H, \lambda_{P, \text{crit}}))
    \label{eq:supp_implicit_critical}
\end{equation}

Due to its implicit form, Eq.~\eqref{eq:supp_implicit_critical} cannot be solved analytically for $\lambda_{P, \text{crit}}$. However, it can be solved numerically for $\lambda_{P, \text{crit}}$ given a set of fixed values for $(H, P, \xi)$, as demonstrated in Fig.~\ref{fig:critical} of the main text.

Similarly, to solve for $\xi_{\text{crit}}$, one can fix the other parameters $(H, P, \lambda_P)$ and find the root of the function $f(\xi_{\text{crit}})$:
\begin{equation}
    f(\xi_{\text{crit}}) = (\lambda_s+\mu)C_2 - (1-\psi)\kappa(\xi_{\text{crit}})(1+C_2) = 0
\label{eq:linear_equation}
\end{equation}
where the function $\kappa(\xi_{\text{crit}})$ is explicitly defined as:
\begin{equation}
    \kappa(\xi_{\text{crit}}) = \beta_0^{\mathrm{STI}}\left((1 - m(H))(1 - P) + (1 - \xi_{\text{crit}} \cdot m(H)) \cdot P\right)
\end{equation}

Thus, Eq.~\eqref{eq:linear_equation} is a linear equation of the form $Y(\xi_\text{crit})=p\cdot\xi_\text{crit} - q$. That allow us to write the explicit solution of $\xi_{crit}$ as below 
\begin{equation}
    \xi_\text{crit}=\frac{q}{p},
\end{equation}
where,
\begin{align}
    q & = (\lambda_s+\mu)C_2+\beta_0^{\mathrm{STI}}(1-\psi)(1+C_2)(m(1-P)-1)\\
    p & = -\beta_0^{\mathrm{STI}}(1-\psi)(1+C_2) \cdot m\cdot P
\end{align}

\subsection{Model Analysis with Influx \texorpdfstring{$\Sigma > 0$}{Sigma > 0}} \label{ssec:supplementary_methods}

This section details the model analysis when considering a constant external influx of infected individuals, $\Sigma > 0$. The fundamental model structure, compartment definitions (susceptible $\Ssti$, asymptomatic infectious $\Iasti$, symptomatic infectious $\Issti$, treated $T$), demographic assumptions (Eq.~\eqref{mu_equal_phi} in Main Text), and parameter definitions (e.g., $\kappa$, $\Gamma_a$, $\Gamma_s$) are identical to those presented in the Main Text. Here, we focus on the modifications to the fixed points and stability criteria due to the presence of influx, presenting minimal calculation steps and final results.

\subsubsection*{Fixed points of the simple HIV-STI model with Influx}
\label{sssec:fixed_points_supp}
The presence of influx ($\Sigma > 0$) modifies the steady-state equations from those presented in the Main Text (Eqs.~\eqref{fpeq:1_main_sigma0}-\eqref{fpeq:3_main_sigma0}). The equations now become:
\begin{align}
0 &= -\kappa \left(I_a^*+I_s^*\right) S^* + \gamma I_a^* + \tilde{\gamma} \left(1 - (S^* + I_a^* + I_s^*)\right) + \mu - \mu S^* - \Sigma \, ,\label{fpeq:1_supp} \\
0 &= \psi \kappa\left(I_a^*+I_s^*\right) S^* - \Gamma_a I_a^* + \psi \Sigma\, , \label{fpeq:2_supp} \\
0 &= (1 - \psi) \kappa\left(I_a^*+I_s^*\right) S^* - \Gamma_s I_s^* + (1 - \psi) \Sigma \label{fpeq:3_supp} \, ,
\end{align}
where $\Gamma_a$ and $\Gamma_s$ are the outflow rates defined in the Main Text (Eqs.~\eqref{seq:asymptomatic_outflow} and \eqref{seq:symptomatic_outflow}).

Despite the addition of $\Sigma$ terms, the algebraic relationship between $I_s^*$ and $I_a^*$ (derived by combining Eqs.~\eqref{fpeq:2_supp} and \eqref{fpeq:3_supp}, noting the cancellation of $\Sigma$-related terms) remains the same as in the no-influx case:
\begin{equation}
I_s^* = C_2 I_a^* \, ,\label{fpeq:Is_Ia_supp_ref}
\end{equation}
where $C_2$ is defined in the Main Text (Eq.~\eqref{fpeq:Is_Ia_main_sigma0}).
The expression for $S^*$ from Eq.~\eqref{fpeq:3_supp} is now:
\begin{equation}
S^* = \frac{ \Gamma_s I_s^* - (1 - \psi) \Sigma }{ (1 - \psi) \kappa (I_a^* + I_s^*) }\, .\label{fpeq:S_supp}
\end{equation}
Substituting Eqs.~\eqref{fpeq:Is_Ia_supp_ref} and \eqref{fpeq:S_supp} into Eq.~\eqref{fpeq:1_supp} leads to a quadratic equation for $I_a^*$:
\begin{equation}
    a(I_a^*)^2 + bI_a^* + c = 0\, ,\label{eq:quadratic_for_Ia_supp}
\end{equation}
where the coefficient $a$ is identical to that in the no-influx case (Eq.~\eqref{fpeq:a_main_sigma0} in Main Text), and $b$ is also identical to its no-influx counterpart (Eq.~\eqref{fpeq:b_main_sigma0} in Main Text). The coefficient $c$, however, is now non-zero due to the influx:
\begin{align}
c &= (\tilde{\gamma} + \mu)(1 - \psi) \Sigma \, .\label{fpeq:coeff_c_supp}
\end{align}

The solution for $I_a^*$ is given by the quadratic formula:
\begin{equation}
    I_a^* = \frac{ -b \pm \sqrt{b^2 - 4ac} }{ 2a }\, . \label{fpeq:Ia_sol_supp}
\end{equation}
Given that $a<0$ and $c>0$ (since $\Sigma>0$), and assuming $b^2-4ac > 0$ for biologically relevant parameter ranges, there exists a unique positive real solution for $I_a^*$. This ensures the existence of a unique endemic equilibrium in the presence of external influx (i.e., $\Sigma > 0$).

\subsubsection*{Calculation of the basic reproduction number in the presence of influx, $R_{0,\Sigma}$}
\label{sssec:Reff_supp}
The basic reproduction number, $R_0$, as derived in the Main Text (Eq.~\eqref{seq:R_0_main}), quantifies the epidemic potential in the absence of influx ($\Sigma=0$) for a fully susceptible population. However, when there is a constant influx of infected individuals ($\Sigma > 0$), as described by the model equations, a true disease-free equilibrium (DFE) does not exist. The disease is perpetually present due to influx.

In this scenario, the critical threshold question is not whether the disease can invade, but whether local transmission can sustain itself and amplify the baseline level of infection. To determine this, we calculate the basic reproduction number in the presence of influx, $R_{0,\Sigma}$, using the next-generation matrix method, evaluated at the endemic equilibrium (EE).

Evaluating $\mathbf{F}$ and $\mathbf{V}$ at the EE ($\Ssti{}^* \neq 0, \Iasti{}^*\neq0, \Issti{}^*\neq0$) and substituting the values in Eq.~\eqref{seq:linearized_infection_subsystem_main}, we get
\begin{align}
F_{\Ssti^*} &=
\begin{pmatrix}
    \psi \kappa \Ssti^* & \psi \kappa \Ssti^* \\
    (1-\psi)\kappa \Ssti^* & (1-\psi)\kappa \Ssti^*
\end{pmatrix} \\
&=
\Ssti^*
\begin{pmatrix}
    \psi \kappa & \psi \kappa \\
    (1-\psi)\kappa & (1-\psi)\kappa
\end{pmatrix}, \\
V &=
\begin{pmatrix}
    \Gamma_a & 0 \\
    0 & \Gamma_s
\end{pmatrix},
\end{align}
where $\Gamma_a$  and $\Gamma_s$ are same as before.
The next-generation matrix is the product $F_{\Ssti^*} V^{-1}$:
\begin{align*}
F_{\Ssti^*} V^{-1} &=
\Ssti^*
\begin{pmatrix}
    \psi \kappa & \psi \kappa \\
    (1-\psi)\kappa & (1-\psi)\kappa
\end{pmatrix}
\begin{pmatrix}
    1/\Gamma_a & 0 \\
    0 & 1/\Gamma_s
\end{pmatrix} \\
&=
\Ssti^*
\begin{pmatrix}
    \psi \kappa / \Gamma_a & \psi \kappa / \Gamma_s \\
    (1-\psi)\kappa / \Gamma_a & (1-\psi)\kappa / \Gamma_s
\end{pmatrix}
\end{align*}
The spectral radius, $\rho(F_{\Ssti^*} V^{-1})$, gives the effective reproduction number. The characteristic equation, $\det(F_{\Ssti^*} V^{-1} - \nu I) = 0$, yields the non-zero eigenvalue:
\begin{align}
    R_{0,\Sigma} = \Ssti^* \left(\frac{\psi \kappa}{\Gamma_a} + \frac{(1-\psi)\kappa}{\Gamma_s}\right)
\end{align}
Recognizing the term in the parenthesis as the original $R_0$, we arrive at the final relationship:
\begin{equation}
    R_{0,\Sigma} = R_0 \cdot \Ssti^*, \label{eq:Reff_supp}
\end{equation}
where $\Ssti^*$ is defined in Eq.~\eqref{fpeq:S_supp}.
This $R_{0,\Sigma}$ serves as a modified threshold for endemic persistence driven by local transmission in the presence of influx.

\subsubsection*{Bifurcation Analysis}
\label{sssec:bifurcation_supp}
With a constant influx of infection ($\Sigma > 0$), the disease cannot be eradicated from the population, meaning $I_a^*$ and $I_s^*$ will not be zero. Even if $R_0 < 1$, infection persists due to external influx. The system no longer has a true DFE. Instead, there is always an endemic equilibrium.
The nature of this endemic equilibrium and its dependence on $R_0$ and $\Sigma$ can be understood through $R_{0,\Sigma} $. The system exhibits behavior analogous to a transcritical bifurcation around $R_{0,\Sigma} =1$.
\begin{itemize}
    \item If $R_{0,\Sigma}  < 1$ the endemic level of infection is primarily sustained by the influx. Local transmission chains may not be self-sustaining.
    \item If $R_{0,\Sigma}  > 1$, local transmission significantly amplifies the infection beyond what influx alone would maintain, leading to a higher, stable endemic prevalence.
\end{itemize}
The influx term $\Sigma$ in Eq.~\eqref{eq:Reff_supp} effectively lowers $R_{0,\Sigma} $ compared to $R_0$, meaning that a higher $R_0$ is required to achieve $R_{0,\Sigma}  > 1$ when influx is present. 

\subsubsection*{Linear Stability Analysis}
\label{sssec:stability_supp}
In the presence of influx ($\Sigma > 0$), a unique endemic equilibrium $(S_{EE}^*, I_{a,EE}^*, I_{s,EE}^*)$ always exists, as determined from Eq.~\eqref{fpeq:Ia_sol_supp}. The stability of this endemic equilibrium is investigated by evaluating the Jacobian matrix $J_{EE}$ at these fixed-point coordinates. The general form of the Jacobian $J$ is identical to that presented in the Main Text (Eq.~\eqref{eq:fulljac_main_sigma0}), but now $S^*, I_a^*, I_s^*$ are the values corresponding to the endemic equilibrium with $\Sigma > 0$.

The characteristic equation remains a cubic polynomial:
\begin{equation}
\nu^3 + c_2 \nu^2 + c_1 \nu + c_0 = 0, \label{eq:char_poly_EE_supp_revised}
\end{equation}
where the coefficients $c_0, c_1, c_2$ are determined from $J_{EE}$ as described in the Main Text. Local asymptotic stability is ensured if the Routh-Hurwitz criteria ($c_2 > 0, c_1 > 0, c_0 > 0,$ and $c_1 c_2 - c_0 > 0$) are met.

For $\Sigma > 0$, the endemic equilibrium is generally found to be stable across relevant parameter ranges, particularly when $R_{0,\Sigma}  > 1$. The influx acts as a constant source of infection, which tends to stabilize the endemic state rather than leading to disease extinction. Detailed algebraic verification of the Routh-Hurwitz conditions for the $\Sigma > 0$ case typically confirms stability when the endemic equilibrium represents a significant level of infection.

Thus, a more detailed sign analysis of the coefficients is made in order to ensure stability across all relevant ranges. We first obtain the specific coefficients. The general Jacobian matrix of this model is:
\[
J = \begin{bmatrix}
-(\Lambda+\mu+\Tilde{\gamma}) & \gamma-(\Tilde{\gamma}+\kappa \Ssti) & -(\Tilde{\gamma} + \kappa \Ssti) \\
\psi \Lambda & \kappa \psi \Ssti - (\gamma+\lambda_a(P,H)+\mu)& \kappa\psi \Ssti \\
(1-\psi)\Lambda & \kappa (1-\psi) \Ssti & \kappa (1-\psi)\Ssti -(\lambda_s(P,H)+\mu)
\end{bmatrix}
\]

We focus here on the endemic equilibrium, which corresponds to a persistent presence of the infection within the population, that is the only feasible positive solution.
In order to study the stability of such solutions, we will plug Eqs.~\eqref{fpeq:Is_Ia_supp_ref} and \eqref{fpeq:S_supp} into the Jacobian and diagonalize it.

We want to find the eigenvalues of the matrix \(\nu J^*\), so we consider the determinant:

\[
\det(J^* - \nu I) = \det\left(
\begin{bmatrix}
-(I_a^* + I_s^*+\mu+\Tilde{\gamma}) & \gamma-(\Tilde{\gamma}+\kappa \Ssti^*) & -(\Tilde{\gamma} + \kappa \Ssti^*) \\
\psi (I_a^* + I_s^*) & \kappa \psi \Ssti^* - (\gamma+\lambda_a+\mu) & \kappa\psi \Ssti^* \\
(1-\psi)(I_a^* + I_s^*) & \kappa (1-\psi) \Ssti^* & \kappa (1-\psi)\Ssti^* -(\lambda_s+\mu)
\end{bmatrix}
- \nu I
\right)
\]

Subtracting \(\nu I\) gives:

\[
=
\det\left(
\begin{bmatrix}
-(I_a^* + I_s^*+\mu+\Tilde{\gamma} + \nu) & \gamma-(\Tilde{\gamma}+\kappa \Ssti^*) & -(\Tilde{\gamma} + \kappa \Ssti^*) \\
\psi (I_a^* + I_s^*) & \kappa \psi \Ssti^* - (\gamma+\lambda_a+\mu + \nu) & \kappa\psi \Ssti^* \\
(1-\psi)(I_a^* + I_s^*) & \kappa (1-\psi) \Ssti^* & \kappa (1-\psi)\Ssti^* -(\lambda_s+\mu + \nu)
\end{bmatrix}
\right)
\]

We now apply the row operation:
\[
R_3 \to R_3 - \frac{1-\psi}{\psi} R_2
\]

which yields the equivalent matrix:

\[
=
\det\left(
\begin{bmatrix}
-(I_a^* + I_s^*+\mu+\Tilde{\gamma} + \nu) & \gamma-(\Tilde{\gamma}+\kappa \Ssti^*) & -(\Tilde{\gamma} + \kappa \Ssti^*) \\
\psi (I_a^* + I_s^*) & \kappa \psi \Ssti^* - (\gamma+\lambda_a+\mu + \nu) & \kappa\psi \Ssti^* \\
0 & \frac{1-\psi}{\psi}(\gamma+\lambda_a+\mu + \nu) & -(\lambda_s+\mu + \nu)
\end{bmatrix}
\right)=0
\]
\[
=
\det\left(
\begin{bmatrix}
a_1 - \nu & a_2 & a_3 \\
a_4 & a_5- \nu & a_6 \\
0 & a_7-\frac{1-\psi}{\psi}\nu& a_8 - \nu
\end{bmatrix}
\right)=(a_1-\nu)\det\left(
\begin{bmatrix}
a_5- \nu & a_6 \\
a_7-\frac{1-\psi}{\psi}\nu& a_8 - \nu
\end{bmatrix}
\right)-a_4\det\left(
\begin{bmatrix}
a_2 & a_3 \\
a_7-\frac{1-\psi}{\psi}\nu& a_8 - \nu
\end{bmatrix}
\right)
\]
\[
=(a_1-\nu)\left[(a_5-\nu)(a_8-\nu)-a_6(a_7-\frac{1-\psi}{\psi}\nu)\right]-a_4\left[a_2(a_8-\nu)-(a_7-\frac{1-\psi}{\psi}\nu)a_3)\right]=0
\]
Expanding the term and writing in the form of \(\nu^3\) polynomial, as in Eq.~\eqref{eq:char_poly_EE_supp_revised}.
Where each coefficient is of the form:
\begin{align}
c_2 &= \frac{1 - \psi}{\psi}\, a_6 - (a_1 + a_5 + a_8), \label{eq:c2}\\[6pt]
c_1 &= 
    a_1 a_5 + a_1 a_8 
    - \frac{1 - \psi}{\psi}\, a_1 a_6 
    - a_6 a_7 + a_5 a_8 
    - a_4 a_2 + \frac{1 - \psi}{\psi}\, a_4 a_3, \label{eq:c1}\\[6pt]
c_0 &= 
    -a_1 a_5 a_8 
    + a_1 a_6 a_7 
    + a_4 a_2 a_8 
    - a_4 a_3 a_7. \label{eq:c0}
\end{align}

The system is stable (Routh-Hurwitz) (i.e., all roots of \(P(\nu)\) have negative real parts if and only if:

\[
\boxed{
\begin{aligned}
&\text{(i) } c_2 > 0 \\
&\text{(ii) } c_0 > 0 \\
&\text{(iii) } c_2 c_1 > c_0
\end{aligned}
}
\]
\begin{itemize}
    \item[(i)] \(c_2 > 0 \;\Rightarrow\;  \dfrac{1 - \psi}{\psi} a_6 - (a_1 + a_5 + a_8) >0 \)
    \end{itemize}
    
    \begin{itemize}
        \item[] where:
        $
        \begin{cases}
            a_1 = -(I_a^* + I_s^* + \mu + \tilde{\gamma}) < 0, & \text{since all constants are positive} \\[2ex]
            a_5 = \kappa \psi S^*-(\gamma + \lambda_a+\mu), & \text{depends on parameters} \\[2ex]
            a_8 = -(\lambda_s+\mu)<0, & \text{since all constants are positive}\\[2ex]
            a_6 = \kappa \psi S^*>0, & \text{so } \dfrac{1 - \psi}{\psi} a_6 >0
        \end{cases}
        $
    \end{itemize}

We see that \(c_2>0\) as long as \(\kappa\psi S^*\) is smaller than the rest of the terms. For the whole spectrum of parameters chosen, \(\kappa\psi S<1\) and the positive terms sum is over 1, hence the positive condition for \(c_2\) is guaranteed.

(ii) \(c_o>0\) if:
\begin{itemize}
        \item[] where:
        $
        \begin{cases}
            a_1 a_5 a_8 < 0, & \text{since \(a_1, a_8<0\) and \(a_5>0\) for the whole spectrum of parameters.} \\[2ex]
            -a_1 a_6 a_7 > 0, & \text{since \(a_1, a_7>0\) and \(a_1<0\)} \\[2ex]
            -a_4 a_2 a_8 < 0, & \text{since \(a_2, a_8<0\) and \(a_4>0\)}\\[2ex]
            a_4 a_3 a_7 < 0 , & \text{since \(a_3<0\) and \(a_4, a_7>0\)}
        \end{cases}
        $
    \end{itemize}

Therefore, as long as the other terms different from \(a_1 a_6 a_7\) are bigger, which holds for the whole scope of parameters used (checked analytically by scanning the phase space of all possible parameters), we will have a positive coefficient \(c_o\).  

(iii) $c_1c_2>c_0$, we can check this by scanning once again our phase space. We found that the coefficients are always larger than $c_0$; therefore, the system is stable due to the Routh-Hurwitz criteria.

\begin{thebibliography}{10}

\bibitem{anderson2012emtricitabine}
Peter~L Anderson, David~V Glidden, Albert Liu, Susan Buchbinder, Javier~R Lama,
  Juan~Vicente Guanira, Vanessa McMahan, Lane~R Bushman, Mart{\'\i}n
  Casap{\'\i}a, Orlando Montoya-Herrera, et~al.
\newblock Emtricitabine-tenofovir concentrations and pre-exposure prophylaxis
  efficacy in men who have sex with men.
\newblock {\em Science translational medicine}, 4(151):151ra125--151ra125,
  2012.

\bibitem{marcus2019risk}
Julia~L Marcus, Kenneth~A Katz, Douglas~S Krakower, and Sarah~K Calabrese.
\newblock Risk compensation and clinical decision making—the case of hiv
  preexposure prophylaxis.
\newblock {\em The New England journal of medicine}, 380(6):510, 2019.

\bibitem{mccormack2016pre}
Sheena McCormack, David~T Dunn, Monica Desai, David~I Dolling, Mitzy Gafos,
  Richard Gilson, Ann~K Sullivan, Amanda Clarke, Iain Reeves, Gabriel Schembri,
  et~al.
\newblock Pre-exposure prophylaxis to prevent the acquisition of hiv-1
  infection (proud): effectiveness results from the pilot phase of a pragmatic
  open-label randomised trial.
\newblock {\em The Lancet}, 387(10013):53--60, 2016.

\bibitem{rozhnova2018elimination}
Ganna Rozhnova, Janneke Heijne, Daniela Bezemer, Ard Van~Sighem, Anne Presanis,
  Daniela De~Angelis, and Mirjam Kretzschmar.
\newblock Elimination prospects of the dutch hiv epidemic among men who have
  sex with men in the era of preexposure prophylaxis.
\newblock {\em Aids}, 32(17):2615--2623, 2018.

\bibitem{murchu2022oral}
Eamon~O Murchu, Liam Marshall, Conor Teljeur, Patricia Harrington, Catherine
  Hayes, Patrick Moran, and Mairin Ryan.
\newblock Oral pre-exposure prophylaxis (prep) to prevent hiv: a systematic
  review and meta-analysis of clinical effectiveness, safety, adherence and
  risk compensation in all populations.
\newblock {\em BMJ open}, 12(5):e048478, 2022.

\bibitem{ayerdi2021low}
Oskar Ayerdi~Aguirrebengoa, Mar Vera~Garc{\'\i}a, Daniel Arias~Ram{\'\i}rez,
  Natalia Gil~Garc{\'\i}a, Teresa Puerta~L{\'o}pez, Petunia Clavo~Escribano,
  Juan Ballesteros~Mart{\'\i}n, Clara Lejarraga~Ca{\~n}as, Nuria
  Fernandez~Pi{\~n}eiro, Manuel~Enrique Fuentes~Ferrer, et~al.
\newblock Low use of condom and high sti incidence among men who have sex with
  men in prep programs.
\newblock {\em PloS one}, 16(2):e0245925, 2021.

\bibitem{milam2019sexual}
Joel Milam, Sonia Jain, Michael~P Dub{\'e}, Eric~S Daar, Xiaoying Sun, Katya
  Corado, Eric Ellorin, Jill Blumenthal, Richard Haubrich, David~J Moore,
  et~al.
\newblock Sexual risk compensation in a pre-exposure prophylaxis demonstration
  study among individuals at risk of hiv.
\newblock {\em JAIDS Journal of Acquired Immune Deficiency Syndromes},
  80(1):e9--e13, 2019.

\bibitem{storholm2017risk}
Erik~D Storholm, Jonathan~E Volk, Julia~L Marcus, Michael~J Silverberg, and
  Derek~D Satre.
\newblock Risk perception, sexual behaviors, and prep adherence among
  substance-using men who have sex with men: a qualitative study.
\newblock {\em Prevention Science}, 18(6):737--747, 2017.

\bibitem{golub2010preexposure}
Sarit~A Golub, William Kowalczyk, Corina~L Weinberger, and Jeffrey~T Parsons.
\newblock Preexposure prophylaxis and predicted condom use among high-risk men
  who have sex with men.
\newblock {\em JAIDS Journal of Acquired Immune Deficiency Syndromes},
  54(5):548--555, 2010.

\bibitem{holt2010gay}
Martin Holt, Diana Bernard, and Kane Race.
\newblock Gay men’s perceptions of sexually transmissible infections and
  their experiences of diagnosis:‘part of the way of life’to feeling
  ‘dirty and ashamed’.
\newblock {\em Sexual Health}, 7(4):411--416, 2010.

\bibitem{hoornenborg2018change}
Elske Hoornenborg, Liza Coyer, Anna van Laarhoven, Roel Achterbergh, Henry
  de~Vries, Maria Prins, Maarten~Schim van~der Loeff, et~al.
\newblock Change in sexual risk behaviour after 6 months of pre-exposure
  prophylaxis use: results from the amsterdam pre-exposure prophylaxis
  demonstration project.
\newblock {\em Aids}, 32(11):1527--1532, 2018.

\bibitem{holt2018community}
Martin Holt, Toby Lea, Limin Mao, Johann Kolstee, Iryna Zablotska, Tim Duck,
  Brent Allan, Michael West, Evelyn Lee, Peter Hull, et~al.
\newblock Community-level changes in condom use and uptake of hiv pre-exposure
  prophylaxis by gay and bisexual men in melbourne and sydney, australia:
  results of repeated behavioural surveillance in 2013--17.
\newblock {\em The lancet HIV}, 5(8):e448--e456, 2018.

\bibitem{quaife2020risk}
Matthew Quaife, Louis MacGregor, Jason~J Ong, Mitzy Gafos, Sergio Torres-Rueda,
  Hannah Grant, Fern Terris-Prestholt, and Peter Vickerman.
\newblock Risk compensation and sti incidence in prep programmes.
\newblock {\em The lancet HIV}, 7(4):e222--e223, 2020.

\bibitem{van2020diverging}
Ward~PH van Bilsen, Anders Boyd, Maarten~FS van~der Loeff, Udi Davidovich,
  Arjan Hogewoning, Lia van~der Hoek, Maria Prins, and Amy Matser.
\newblock Diverging trends in incidence of hiv versus other sexually
  transmitted infections in hiv-negative msm in amsterdam.
\newblock {\em Aids}, 34(2):301--309, 2020.

\bibitem{spinner2019summary}
CD~Spinner, GF~Lang, C~Boesecke, H~Jessen, K~Schewe, German-Austrian~PrEP
  consensus conference~meeting 2018, Hans-J{\"u}rgen Stellbrink, Stefan Esser,
  Annette Haberl, Katja R{\"o}mer, et~al.
\newblock Summary of german-austrian hiv prep guideline.
\newblock {\em HIV medicine}, 20(6):368--376, 2019.

\bibitem{contreras2021challenges}
Sebastian Contreras, Jonas Dehning, Matthias Loidolt, Johannes Zierenberg,
  F~Paul Spitzner, Jorge~H Urrea-Quintero, Sebastian~B Mohr, Michael Wilczek,
  Michael Wibral, and Viola Priesemann.
\newblock The challenges of containing {SARS-CoV-2} via test-trace-and-isolate.
\newblock {\em Nature communications}, 12(1):1--13, 2021.

\bibitem{wagner2025societal}
Joel Wagner, Simon Bauer, Sebastian Contreras, Luk Fleddermann, Ulrich Parlitz,
  and Viola Priesemann.
\newblock Societal self-regulation induces complex infection dynamics and
  chaos.
\newblock {\em Physical Review Research}, 7(1):013308, 2025.

\bibitem{donges2022interplay}
Philipp D{\"o}nges, Joel Wagner, Sebastian Contreras, Emil~N Iftekhar, Simon
  Bauer, Sebastian~B Mohr, Jonas Dehning, Andr{\'e} Calero~Valdez, Mirjam
  Kretzschmar, Michael M{\"a}s, et~al.
\newblock Interplay between risk perception, behavior, and covid-19 spread.
\newblock {\em Frontiers in Physics}, 10:842180, 2022.

\bibitem{testing_paradox}
Laura Müller, Piklu Mallick, Antonio~B. Marín-Carballo, Philipp Dönges,
  Robyn J.~N. Kettlitz, Carolina~J. Klett-Tammen, Mirjam Kretzschmar, Viola
  Priesemann, and Seba Contreras.
\newblock Testing paradox may explain increased observed prevalence of
  bacterial stis among msm on hiv prep: A modeling study.
\newblock {\em Proceedings of the National Academy of Sciences},
  122(44):e2524944122, 2025.

\bibitem{walker2022assessing}
Michael~L Walker, David Stiasny, Rebecca~J Guy, Matthew~G Law, Martin Holt,
  Limin Mao, Basil Donovan, Andrew~E Grulich, Richard~T Gray, and David~G
  Regan.
\newblock Assessing the impact of hiv preexposure prophylaxis scale-up on
  gonorrhea incidence among gay and bisexual men in sydney: a mathematical
  modeling study.
\newblock {\em Sexually Transmitted Diseases}, 49(8):534--540, 2022.

\bibitem{jenness2017incidence}
Samuel~M Jenness, Kevin~M Weiss, Steven~M Goodreau, Thomas Gift, Harrell
  Chesson, Karen~W Hoover, Dawn~K Smith, Albert~Y Liu, Patrick~S Sullivan, and
  Eli~S Rosenberg.
\newblock Incidence of gonorrhea and chlamydia following human immunodeficiency
  virus preexposure prophylaxis among men who have sex with men: a modeling
  study.
\newblock {\em Clinical Infectious Diseases}, 65(5):712--718, 2017.

\bibitem{pharaon_impact_2020}
Joe Pharaon and Chris~T. Bauch.
\newblock The {Impact} of {Pre}-exposure {Prophylaxis} for {Human}
  {Immunodeficiency} {Virus} on {Gonorrhea} {Prevalence}.
\newblock {\em Bulletin of Mathematical Biology}, 82(7):85, July 2020.

\bibitem{escobar_agent-based_2015}
Erik Escobar, Ryan Durgham, Olaf Dammann, and Thomas Stopka.
\newblock Agent-based computational model of the prevalence of gonococcal
  infections after the implementation of {HIV} pre-exposure prophylaxis
  guidelines.
\newblock {\em Online Journal of Public Health Informatics}, 7(3):e61710,
  December 2015.
\newblock Company: Online Journal of Public Health Informatics Distributor:
  Online Journal of Public Health Informatics Institution: Online Journal of
  Public Health Informatics Label: Online Journal of Public Health Informatics
  Publisher: JMIR Publications Inc., Toronto, Canada.

\bibitem{abiodun2022mathematical}
Oluwakemi~E Abiodun, Olukayode Adebimpe, James~A Ndako, Olajumoke Oludoun,
  Benedicta Aladeitan, and Michael Adeniyi.
\newblock Mathematical modeling of hiv-hcv co-infection model: Impact of
  parameters on reproduction number.
\newblock {\em F1000Research}, 11:1153, 2022.

\bibitem{saha2021global}
Pritam Saha and Uttam Ghosh.
\newblock Global dynamics and control strategies of an epidemic model having
  logistic growth, non-monotone incidence with the impact of limited hospital
  beds.
\newblock {\em Nonlinear Dynamics}, 105(1):971--996, 2021.

\bibitem{jing2024backward}
Shuanglin Jing, Ling Xue, and Jichen Yang.
\newblock Backward bifurcation arising from decline of immunity against
  emerging infectious diseases.
\newblock {\em Applied Mathematics Letters}, 158:109241, 2024.

\bibitem{wang2022bifurcations}
Shaoli Wang, Tengfei Wang, and Yuming Chen.
\newblock Bifurcations and bistability of an age-structured viral infection
  model with a nonmonotonic immune response.
\newblock {\em International Journal of Bifurcation and Chaos}, 32(10):2250151,
  2022.

\bibitem{saha2025dynamic}
Pritam Saha, Kalyan Kumar~Pal, Uttam Ghosh, and Pankaj Kumar~Tiwari.
\newblock Dynamic analysis of deterministic and stochastic seir models
  incorporating the ornstein--uhlenbeck process.
\newblock {\em Chaos: An Interdisciplinary Journal of Nonlinear Science},
  35(2), 2025.

\bibitem{pradeesh2025bistability}
M~Pradeesh and Prakash Mani.
\newblock Bistability and bifurcations in hiv-1 infection model with
  non-monotone responses.
\newblock {\em Scientific Reports}, 15(1):7265, 2025.

\bibitem{corcoran2021low}
Carl Corcoran and Alan Hastings.
\newblock A low-dimensional network model for an sis epidemic: analysis of the
  super compact pairwise model.
\newblock {\em Bulletin of Mathematical Biology}, 83(7):77, 2021.

\bibitem{montes2022evaluating}
Sandra Montes-Olivas, Yaz Ozten, Martin Homer, Katy Turner, Christopher~K
  Fairley, Jane~S Hocking, Desiree Tse, Nicolas Verschueren~van Rees,
  William~CW Wong, and Jason~J Ong.
\newblock Evaluating the impact and cost-effectiveness of chlamydia management
  strategies in hong kong: A modeling study.
\newblock {\em Frontiers in Public Health}, 10:932096, 2022.

\bibitem{qu2021effect}
Zhuolin Qu, Asma Azizi, Norine Schmidt, Megan~Clare Craig-Kuhn, Charles
  Stoecker, James Mac~Hyman, and Patricia~J Kissinger.
\newblock Effect of screening young men for chlamydia trachomatis on the rates
  among women: a network modelling study for high-prevalence communities.
\newblock {\em BMJ open}, 11(1):e040789, 2021.

\bibitem{world2021consolidated}
World~Health Organization et~al.
\newblock {\em Consolidated guidelines on HIV prevention, testing, treatment,
  service delivery and monitoring: recommendations for a public health
  approach}.
\newblock World Health Organization, 2021.

\bibitem{hevey2018prep}
Matthew~A Hevey, Jennifer~L Walsh, and Andrew~E Petroll.
\newblock Prep continuation, hiv and sti testing rates, and delivery of
  preventive care in a clinic-based cohort.
\newblock {\em AIDS Education and Prevention}, 30(5):393--405, 2018.

\bibitem{lamontagne2003chlamydia}
D~Scott LaMontagne, David~N Fine, and Jeanne~M Marrazzo.
\newblock Chlamydia trachomatis infection in asymptomatic men.
\newblock {\em American journal of preventive medicine}, 24(1):36--42, 2003.

\bibitem{korenromp2002proportion}
Eline~L Korenromp, Mondastri~K Sudaryo, Sake~J de~Vlas, Ronald~H Gray, Nelson~K
  Sewankambo, David Serwadda, Maria~J Wawer, and J~Dik~F Habbema.
\newblock What proportion of episodes of gonorrhoea and chlamydia becomes
  symptomatic?
\newblock {\em International journal of STD \& AIDS}, 13(2):91--101, 2002.

\bibitem{kesler2016actual}
Maya~A Kesler, Rupert Kaul, Juan Liu, Mona Loutfy, Dionne Gesink, Ted Myers,
  and Robert~S Remis.
\newblock Actual sexual risk and perceived risk of hiv acquisition among
  hiv-negative men who have sex with men in toronto, canada.
\newblock {\em BMC public health}, 16:1--9, 2016.

\bibitem{nagumo1942lage}
Mitio Nagumo.
\newblock {\"U}ber die lage der integralkurven gew{\"o}hnlicher
  differentialgleichungen.
\newblock {\em Proceedings of the physico-mathematical society of Japan. 3rd
  Series}, 24:551--559, 1942.

\bibitem{diekmann2010next}
Odo Diekmann, Johan Andre~Peter Heesterbeek, and Michael~G Roberts.
\newblock The construction of next-generation matrices for compartmental
  epidemic models.
\newblock {\em Journal of the royal society interface}, 7(47):873--885, 2010.

\bibitem{jiao2023mpox}
Kedi Jiao, Yutong Xu, Siwen Huang, Yuhang Zhang, Jingtao Zhou, Yan Li, Yongkang
  Xiao, Wei Ma, Lin He, Xianlong Ren, et~al.
\newblock Mpox risk perception and associated factors among chinese young men
  who have sex with men: Results from a large cross-sectional survey.
\newblock {\em Journal of Medical Virology}, 95(8):e29057, 2023.

\bibitem{khawcharoenporn2019hiv}
Thana Khawcharoenporn, Suteera Mongkolkaewsub, Chanon Naijitra, Worawoot
  Khonphiern, Anucha Apisarnthanarak, and Nittaya Phanuphak.
\newblock Hiv risk, risk perception and uptake of hiv testing and counseling
  among youth men who have sex with men attending a gay sauna.
\newblock {\em AIDS research and therapy}, 16(1):13, 2019.

\bibitem{balan2019low}
Iv{\'a}n~C Bal{\'a}n, Javier Lopez-Rios, Curtis Dolezal, Christine~Tagliaferri
  Rael, and Cody Lentz.
\newblock Low sexually transmissible infection knowledge, risk perception and
  concern about infection among men who have sex with men and transgender women
  at high risk of infection.
\newblock {\em Sexual health}, 16(6):580--586, 2019.

\bibitem{clutterbuck20182016}
Dan Clutterbuck, David Asboe, Tristan Barber, Carol Emerson, Nigel Field,
  Stuart Gibson, Gwenda Hughes, Rachael Jones, Martin Murchie, Achyuta~V Nori,
  et~al.
\newblock 2016 united kingdom national guideline on the sexual health care of
  men who have sex with men.
\newblock {\em International journal of STD \& AIDS}, page 0956462417746897,
  2018.

\bibitem{centers2018preexposure}
{United States. Public Health Service.}, {Centers for Disease Control and
  Prevention (U.S.)}, and {National Center for HIV/AIDS, Viral Hepatitis, STD,
  and TB Prevention (U.S.). Division of HIV/AIDS Prevention. Preexposure
  Prophylaxis Work Group.}
\newblock Preexposure prophylaxis for the prevention of hiv infection in the
  united states – 2021 update clinical providers’ supplement, December
  2021.
\newblock \url{https://stacks.cdc.gov/view/cdc/112359} (accessed November 12,
  2025).

\bibitem{deutsche_aidshilfe_prepchecks}
{Deutsche Aidshilfe}.
\newblock Prep-checks und medizinische begleitung, 2025.
\newblock \url{https://www.aidshilfe.de/de/hiv-prep/prep-checks} (accessed
  September 3, 2025).

\bibitem{van2024belgian}
Jens~T Van~Praet, Sophie Henrard, Chris Kenyon, Agn{\`e}s Libois, Annelies
  Meuwissen, Anne-Sophie Sauvage, Anne Vincent, Jef Vanhamel, Gert Scheerder,
  Belgian~Research on~AIDS, and HIV~Consortium (BREACH).
\newblock Belgian 2024 guidance on the use of pre-exposure prophylaxis.
\newblock {\em Acta Clinica Belgica}, 79(2):121--129, 2024.

\bibitem{white20252025}
John~A White, Nicole~HTM Dukers-Muijrers, Christian~JPA Hoebe, Chris~R Kenyon,
  Jonathan Dc~Ross, and Magnus Unemo.
\newblock 2025 european guideline on the management of chlamydia trachomatis
  infections.
\newblock {\em International journal of STD \& AIDS}, 36(6):434--449, 2025.

\bibitem{kenyon2023management}
Chris Kenyon, Bj{\"o}rn Herrmann, Gwenda Hughes, and Henry~JC de~Vries.
\newblock Management of asymptomatic sexually transmitted infections in europe:
  towards a differentiated, evidence-based approach.
\newblock {\em The Lancet Regional Health--Europe}, 34, 2023.

\bibitem{williams2023frequent}
Eloise Williams, Deborah~A Williamson, and Jane~S Hocking.
\newblock Frequent screening for asymptomatic chlamydia and gonorrhoea
  infections in men who have sex with men: time to re-evaluate?
\newblock {\em The Lancet Infectious Diseases}, 23(12):e558--e566, 2023.

\bibitem{schnyder2025understanding}
Simon~K Schnyder, John~J Molina, Ryoichi Yamamoto, and Matthew~S Turner.
\newblock Understanding nash epidemics.
\newblock {\em Proceedings of the National Academy of Sciences},
  122(9):e2409362122, 2025.

\bibitem{kettlitz2023association}
R~Kettlitz, M~Harries, J~Ortmann, G~Krause, A~Aigner, and B~Lange.
\newblock Association of known sars-cov-2 serostatus and adherence to personal
  protection measures and the impact of personal protective measures on
  seropositivity in a population-based cross-sectional study (muspad) in
  germany.
\newblock {\em BMC Public Health}, 23(1):2281, 2023.

\bibitem{ViiVHealthcare2022EMAcabotegravir}
{ViiV Healthcare}.
\newblock European medicines agency validates viiv healthcare's marketing
  authorisation application for cabotegravir long-acting injectable for hiv
  prevention, October 2022.
\newblock Press Release. Accessed: 2025-11-12.

\bibitem{bekker2024twice}
Linda-Gail Bekker, Moupali Das, Quarraisha Abdool~Karim, Khatija Ahmed, Joanne
  Batting, William Brumskine, Katherine Gill, Ishana Harkoo, Manjeetha
  Jaggernath, Godfrey Kigozi, et~al.
\newblock Twice-yearly lenacapavir or daily f/taf for hiv prevention in
  cisgender women.
\newblock {\em New England Journal of Medicine}, 391(13):1179--1192, 2024.

\bibitem{hillier2021safety}
S~Hillier, LG~Bekker, SA~Riddler, CW~Hendrix, S~Badal-Faesen, S~Rasmussen,
  H~Schwartz, P~Macdonald, J~Lombaard, Y~Caraco, et~al.
\newblock Safety and pharmacokinetics of oral islatravir once monthly for hiv
  pre-exposure prophylaxis (prep): week 24 analysis of a phase 2a trial.
\newblock {\em Journal of the International AIDS Society}, 24(S4):72--74, 2021.

\bibitem{wang2025oral}
Haoyi Wang, Alejandro~Adriaque Lozano, Johann Kolstee, Hanne~ML Zimmermann,
  Jonathan Tosh, Melanie Schroeder, Ama Appiah, and Kai~J Jonas.
\newblock Oral prep use and intention to use long-acting prep regimens among
  msm accessing prep via governmental and non-governmental provision pathways,
  20 european countries, october 2023 to april 2024.
\newblock {\em Eurosurveillance}, 30(34):2500122, 2025.

\bibitem{rozhnova2016impact}
Ganna Rozhnova, Maarten F~Schim van~der Loeff, Janneke~CM Heijne, and Mirjam~E
  Kretzschmar.
\newblock Impact of heterogeneity in sexual behavior on effectiveness in
  reducing hiv transmission with test-and-treat strategy.
\newblock {\em PLoS computational biology}, 12(8):e1005012, 2016.

\bibitem{khosropour2021changing}
Christine~M Khosropour, Julia~C Dombrowski, Lindley~A Barbee, Roxanne~P Kerani,
  Anna Berzkalns, and Matthew~R Golden.
\newblock Changing patterns of sexual behavior and hiv/sti among men who have
  sex with men in seattle, 2002 to 2018.
\newblock {\em JAIDS Journal of Acquired Immune Deficiency Syndromes},
  87(4):1032--1039, 2021.

\bibitem{hess2017trends}
Kristen~L Hess, Nicole Crepaz, Charles Rose, David Purcell, and Gabriela
  Paz-Bailey.
\newblock Trends in sexual behavior among men who have sex with men (msm) in
  high-income countries, 1990--2013: a systematic review.
\newblock {\em AIDS and Behavior}, 21(10):2811--2834, 2017.

\bibitem{sandfort2013sexual}
Theo Sandfort, Huso Yi, Justin Knox, and Vasu Reddy.
\newblock Sexual partnership types as determinant of hiv risk in south african
  msm: an event-level cluster analysis.
\newblock {\em AIDS and Behavior}, 17(Suppl 1):23--32, 2013.

\bibitem{hoff2012relationship}
Colleen~C Hoff, Deepalika Chakravarty, Sean~C Beougher, Torsten~B Neilands, and
  Lynae~A Darbes.
\newblock Relationship characteristics associated with sexual risk behavior
  among msm in committed relationships.
\newblock {\em AIDS patient care and STDs}, 26(12):738--745, 2012.

\bibitem{tien2020antimicrobial}
Vivian Tien, Chitra Punjabi, and Marisa~K Holubar.
\newblock Antimicrobial resistance in sexually transmitted infections.
\newblock {\em Journal of travel medicine}, 27(1):taz101, 2020.

\bibitem{de2022worryingly}
Irith De~Baetselier, Bea Vuylsteke, Thijs Reyniers, Hilde Smet, Dorien Van~den
  Bossche, Chris Kenyon, and Tania Crucitti.
\newblock Worryingly high prevalence of resistance-associated mutations to
  macrolides and fluoroquinolones in mycoplasma genitalium among men who have
  sex with men with recurrent sexually transmitted infections.
\newblock {\em International journal of STD \& AIDS}, 33(4):385--390, 2022.

\bibitem{wilting2018inferring}
Jens Wilting and Viola Priesemann.
\newblock Inferring collective dynamical states from widely unobserved systems.
\newblock {\em Nature communications}, 9(1):2325, 2018.

\bibitem{levina2017subsampling}
Anna Levina and Viola Priesemann.
\newblock Subsampling scaling.
\newblock {\em Nature communications}, 8(1):15140, 2017.

\bibitem{levina2022tackling}
Anna Levina, Viola Priesemann, and Johannes Zierenberg.
\newblock Tackling the subsampling problem to infer collective properties from
  limited data.
\newblock {\em Nature Reviews Physics}, 4(12):770--784, 2022.

\end{thebibliography}
\end{document}